\documentclass[twocolumn, twocolappendix]{aastex631} 
\pdfoutput=1

\usepackage{hyperref}
\usepackage{graphicx}
\usepackage{gensymb}
\usepackage{amsmath}
\usepackage{float}
\usepackage{threeparttable}
\usepackage{amssymb}
\usepackage{mathrsfs}
\usepackage{afterpage}
\usepackage{scalerel}
\usepackage{natbib}
\bibpunct[; ]{(}{)}{;}{a}{}{,}
\usepackage{multirow}
\usepackage{graphics}
\usepackage{threeparttable}
\usepackage[varg]{txfonts}
\usepackage{xcolor}
\usepackage{bm}
\usepackage{wrapfig}
\usepackage{mathtools}
\usepackage[mathscr]{euscript}
\makeatletter
\newcommand*{\rom}[1]{\expandafter\@slowromancap\romannumeral #1@}
\makeatother
\usepackage[mathscr]{euscript}

\DeclareSymbolFont{rsfs}{U}{rsfs}{m}{n}
\DeclareSymbolFontAlphabet{\mathscrsfs}{rsfs}

\begin{document}

\title{A Rapidly Accreting Active Galactic Nucleus Hidden in a Dust-Obscured Galaxy at $z \sim 0.8$}

\author[0000-0001-6317-8488]{Nathan Cristello$^\star$}\let\thefootnote\relax\footnote{$^\star$Email: \href{nathancristello@gmail.com}{nathancristello@gmail.com}}
\affiliation{Department of Astronomy and Astrophysics, 525 Davey Lab, The Pennsylvania State University, University Park, PA 16802, USA}

\author[0000-0002-4436-6923]{Fan Zou}
\affiliation{Department of Astronomy, University of Michigan, Ann Arbor, MI 48109, USA}
\affiliation{Department of Astronomy and Astrophysics, 525 Davey Lab, The Pennsylvania State University, University Park, PA 16802, USA}
\affiliation{Institute for Gravitation and the Cosmos, The Pennsylvania State University, University Park, PA 16802, USA}

\author[0000-0002-0167-2453]{William N. Brandt}
\affiliation{Department of Astronomy and Astrophysics, 525 Davey Lab, The Pennsylvania State University, University Park, PA 16802, USA}
\affiliation{Institute for Gravitation and the Cosmos, The Pennsylvania State University, University Park, PA 16802, USA}
\affiliation{Department of Physics, 104 Davey Laboratory, The Pennsylvania State University, University Park, PA 16802, USA}

\author[0000-0002-6990-9058]{Zhibo Yu}
\affiliation{Department of Astronomy and Astrophysics, 525 Davey Lab, The Pennsylvania State University, University Park, PA 16802, USA}
\affiliation{Institute for Gravitation and the Cosmos, The Pennsylvania State University, University Park, PA 16802, USA}

\author[0000-0003-0680-9305]{Fabio Vito}
\affiliation{INAF – Osservatorio di Astrofisica e Scienza dello Spazio di Bologna, Via Gobetti 93/3, I-40129 Bologna, Italy}

\author[0000-0002-1653-4969]{Shifu Zhu}
\affiliation{CAS Key Laboratory for Research in Galaxies and Cosmology, Department of Astronomy, University of Science and Technology of China, Hefei 230026, China}
\affiliation{School of Astronomy and Space Sciences, University of Science and Technology of China, Hefei 230026, China}

\begin{abstract}
Dust-obscured galaxies (DOGs) containing central supermassive black holes (SMBHs) that are rapidly accreting (i.e., having high Eddington ratios, $\lambda_\mathrm{Edd}$) may represent a key phase closest to the peak of both the black-hole and galaxy growth in the coevolution framework for SMBHs and galaxies. In this work, we present a 68 ks XMM-Newton observation of the high-$\lambda_\mathrm{Edd}$ DOG J1324+4501 at \mbox{$z \sim 0.8$}, which was initially observed by Chandra. We analyze the XMM-Newton spectra jointly with archival Chandra spectra. In performing a detailed \mbox{X-ray} spectral analysis, we find that the source is intrinsically \mbox{X-ray} luminous with $\log (L_\mathrm{X}$/erg s$^{-1}) = 44.71^{+0.08}_{-0.12}$ and heavily obscured with $\log (N_\mathrm{H}/\mathrm{cm}^{-2}) = 23.43^{+0.09}_{-0.13}$. We further utilize UV-to-IR archival photometry to measure and fit the source's spectral energy distribution (SED) to estimate its host-galaxy properties. We present a supplementary comparison sample of 21 \mbox{X-ray} luminous DOGs from the XMM-SERVS survey with sufficient ($> 200$) $0.5-10$ keV counts to perform a similarly detailed X-ray spectral analysis. Of the X-ray luminous DOGs in our sample, we find that J1324+4501 is the most remarkable, possessing one of the highest \mbox{X-ray} luminosities, column densities, and star-formation rates. We demonstrate that J1324+4501 is in an extreme evolutionary stage where SMBH accretion and galaxy growth are at their peaks.
\end{abstract}


\section{Introduction} \label{sec:intro}
Under the coevolution framework for supermassive black holes (SMBHs) and their host galaxies (e.g., \citealt{Hopkins+06, Hopkins+08}; \citealt{Alexander+12}), the peak of both SMBH accretion and star formation occurs during dust-enshrouded, heavily obscured phases following mergers among gas-rich galaxies. During its early stage, a large amount of material fuels the obscured SMBH with accretion approaching the Eddington limit; then, radiation-driven outflows from near the central SMBH sweep out the obscuring material, allowing the SMBH to shine as an unobscured quasar (e.g., \citealt{Glikman+12}; \citealt{Brusa+15}).

The successes of the Spitzer Space Telescope (\citealt{Werner+04}) and the Wide-field Infrared Survey Explorer (WISE; \citealt{Wright+10}) have enabled detailed analyses of dust-obscured galaxies (DOGs) and Hot DOGs. DOGs and Hot DOGs, observationally selected via their extremely red optical-to-infrared (IR) colors (e.g., \citealt{Dey+08}; \citealt{Wu+12}; \citealt{Toba+17}), may often represent the peak phase in the SMBH-host galaxy coevolution framework.

In particular, Hot DOGs, which are a rare sub-population of DOGs with a sky surface density of around one candidate per 30 deg$^2$, are believed to be mainly powered by deeply buried, massive, and rapidly accreting SMBHs with high Eddington ratios ($\lambda_\mathrm{Edd}$). Their extreme IR colors arise from hot dust emission heated by the central accreting SMBH, with dust temperatures reaching hundreds of K (e.g., \citealt{Tsai+15}). They often have high intrinsic rest-frame $2-10$ keV luminosities ($L_\mathrm{X}$) with nearly Compton-thick (CT) obscuration (e.g., \citealt{Vito+18}).

DOGs are generally less extreme with less-massive SMBHs, larger host-galaxy contributions, smaller dust temperatures ($30-40$ K), and they generally have smaller $L_\mathrm{X}$ than Hot DOGs. The DOG population is heterogeneous – they appear to span a wide range of evolutionary stages or even often can be explained by episodes of star formation (e.g., \citealt{Lanzuisi+09}; \citealt{Corral+16}; \citealt{Yu+24}). However, high-$\lambda_\mathrm{Edd}$ DOGs are thought to be analogous to Hot DOGs at the peak evolutionary stage following gas-rich mergers (\citealt{Zou+20}). The \mbox{X-ray} obscuration of DOGs spans a wide range from low-to-moderate to high column densities ($N_\mathrm{H}$), and high-$\lambda_\mathrm{Edd}$ DOGs are expected to have the generally highest $N_\mathrm{H}$. Due partly to the relatively smaller $L_\mathrm{X}$ of DOGs compared to Hot DOGs, high-$\lambda_\mathrm{Edd}$ DOGs have not been well sampled in the X-ray regime. \cite{Zou+20} conducted systematic Chandra snapshot ($3-5$ ks per source) observations of 12 high-$\lambda_\mathrm{Edd}$ DOGs, but most sources are either undetected or have very limited ($\le$ 2) counts. 

Previous \mbox{X-ray} observations of high-$\lambda_\mathrm{Edd}$ (Hot) DOGs are highly limited by the source counts. There is only one source, W1835+4355, with X-ray net counts above 130, regardless of the energy band (i.e., either from Chandra, XMM-Newton, or NuSTAR), and W1835+4355 itself has 177 net counts in the $0.5-8$~keV band from a 42~ks XMM-Newton observation and 61 net counts in the $3-24$~keV band from a 155~ks NuSTAR observation (\citealt{Piconcelli+15}; \citealt{Zappacosta+18}). W1835+4355 was the first Hot DOG with moderately detailed spectral analyses presented; its \mbox{X-ray} spectrum shows a prominent neutral iron (Fe) K$\alpha$ line, a tentative ionized Fe line, and a strong scattered soft component. Aside from this single Hot DOG, most of the other high-$\lambda_\mathrm{Edd}$ (Hot) DOGs lack sufficient counts to examine more detailed spectral features other than estimating $N_\mathrm{H}$ and $L_\mathrm{X}$, and many strong assumptions (e.g., fixing the photon index) were often made when estimating $N_\mathrm{H}$, $L_\mathrm{X}$, and occasionally Fe K$\alpha$ lines (e.g., \citealt{Vito+18}). This is not because high-$\lambda_\mathrm{Edd}$ (Hot) DOGs lack \mbox{X-ray} observations; instead, there are around ten such sources with more than 50 ks of exposure each from Chandra, XMM-Newton, and/or NuSTAR, and around two dozen sources with shorter exposures. However, spectral analyses of high-$\lambda_\mathrm{Edd}$ (Hot) DOGs have been hindered by their low \mbox{X-ray} count rates.

Fortunately, \cite{Zou+20} found a uniquely \mbox{X-ray} bright high-$\lambda_\mathrm{Edd}$ DOG, SDSS J132440.17+450133.8 (J1324+4501 hereafter), with a 3 ks Chandra snapshot observation. J1324+4501 has $\lambda_\mathrm{Edd}=1.13_{-0.71}^{+1.34}$ (see Section \ref{sec:hosts}) and $z_\mathrm{spec}=0.774$. It has an even higher $2-10$ keV \mbox{X-ray} flux than W1835+4355 ($2\times10^{-13}$ erg cm$^{-2}$ s$^{-1}$), and thus is the brightest source among high-$\lambda_\mathrm{Edd}$ (Hot) DOGs, offering us a unique opportunity to examine the detailed \mbox{X-ray} spectrum of this extreme population. In this work, we obtained 68 ks follow-up XMM-Newton observations for J1324+4501, aiming to have a much better \mbox{X-ray} characterization of it compared to the previous \mbox{X-ray} snapshot. We perform a broad-band, $\sim$0.5–10 keV joint \mbox{XMM-Newton/Chandra} spectral analysis and updated \mbox{X-ray}-to-mid-IR spectral energy distribution (SED) modeling. 

To support our detailed analysis of J1324+4501, we also present the \mbox{X-ray} spectral analysis and SED results for 21 other typical, \mbox{X-ray} bright ($>$ 200 counts) DOGs, without the high-$\lambda_\mathrm{Edd}$ requirement, residing in the XMM-Spitzer Extragalactic Representative Volume Survey (XMM-SERVS; \citealt{Chen+18}; \citealt{Ni+21_xmmservs}). We acknowledge that, in selecting these DOGs with the highest counts, we are biasing our analysis toward those with high \mbox{X-ray} luminosities and low column densities. However, these high counts are needed in order to perform a meaningful spectral analysis. These sources provide us with a diverse sample of DOGs across a wide range of redshift ($z \approx 0.98-3.0$) that allows for a robust comparison between J1324+4501 and similarly \mbox{X-ray} luminous, but physically different, DOGs. Because previous \mbox{X-ray} analyses of DOGs have been greatly hindered by low-quality \mbox{X-ray} spectra, our analysis across these three fields provides an unprecedented look into the \mbox{X-ray} properties of DOGs. 

The structure of this paper is as follows. In Section \ref{sec:data}, we describe the XMM-Newton \mbox{X-ray} observations, the available archival Chandra data, and the data-reduction methods used. In Section \ref{sec:modeling}, we outline our \mbox{X-ray} spectral fitting process. In Section \ref{sec:hosts}, we obtain SED measurements for each of the sources in our sample. In Section \ref{sec:results}, we display the results of our multiwavelength analysis and compare our DOG to previously reported (Hot) DOGs from the literature. Lastly, Section \ref{sec:disc} summarizes this work. Throughout this paper, we adopt a flat $\Lambda$CDM cosmology with $H_0$ = 70 km s$^{-1}$ Mpc$^{-1}$, $\Omega_\Lambda$ = 0.7, and $\Omega_\mathrm{M}$ = 0.3, and uncertainties are reported at the $1\sigma$ level unless otherwise noted.

\section{X-ray Observations and Data Reduction} \label{sec:data}
In this section, we detail the \mbox{X-ray} observations and data reduction processes for J1324+4501 and our supplementary sample of 21 other \mbox{X-ray} luminous DOGs. 

\subsection{J1324+4501} \label{sec:j1324_red}
J1324+4501 was one of 36 IR-bright DOGs studied in \cite{Toba+17}, who selected these based upon their extreme optical/IR colors and clear [O \rom{3}] emission lines in their SDSS spectra. J1324+4501 was further observed by Chandra (PI Garmire; ObsID 21144) for 3.1 ks on August 22, 2019, and the results were reported in \cite{Zou+20}. It was clearly detected with 15 counts between $2-7$ keV but 0 counts below 2 keV. This hard spectrum indicates likely heavy intrinsic obscuration. 

J1324+4501 was further observed with XMM-Newton (\citealt{Jansen+01}) for 67.9 ks (PI Zou; ObsID 0921650101) on December 18, 2023. Due to high levels of background flaring during the observation, the MOS observations were broken into 8 exposures. These EPIC observations were performed with the PN and MOS cameras operating in Full-Window mode with the Thin filter applied. We reduce these observations and extract the corresponding spectra using the XMM-Newton Science Analysis System (SAS) v21.0.0. We filter the \mbox{X-ray} event lists to ignore periods of high background flaring activity by selecting good time intervals with the time intervals exceeding count rates 3$\sigma$ above the mean count rate value being removed. We obtain 34.2, 32.3, and 14.4 ks of flare-filtered exposure for the MOS1, MOS2, and PN cameras, respectively. We extract the source spectrum using a circular cell with 20.0" radius centered on the position of the source, and we extract the background using an 85" radius circular source-free region elsewhere on the same CCD chip\footnote{\textsuperscript{1}A 20" radius corresponds to an enclosed energy fraction of 70-80\%, and we verify that our results are materially unaffected by choice of radius.}.\textsuperscript{1} We further merge the spectra belonging to the same camera but from different exposures into a single one using the SAS task \texttt{epicspeccombine}. Thus, we are left with three spectra from EPIC MOS1, MOS2, and PN, and we fit them jointly rather than merging them into one spectrum. Lastly, we group these spectra and their background spectra to at least one count per bin for spectral inference.

For our XMM-Newton observation, J1324+4501 is detected in the full ($0.5-10$ keV), soft ($0.5-2$ keV), and hard ($2-10$ keV) bands. We obtained 468 (270), 124 (46), and 344 (224) total (net), aperture-uncorrected counts in the full, soft, and hard bands, respectively, for all the EPIC cameras combined. 

\subsection{Supplementary Comparison Sample}
The additional 21 sources in our study were observed as part of the XMM-SERVS survey. The XMM-SERVS survey is a $\sim50$ ks depth \mbox{\mbox{X-ray}} survey that covers the prime parts of three out of the five Vera C. Rubin Observatory Legacy Survey of Space and Time Deep-Drilling Fields (LSST DDFs): W-CDF-S (Wide Chandra Deep Field-South; 4.6 deg$^2$), ELAIS-S1 (European Large-Area ISO Survey-S1; 3.2 deg$^2$), and XMM-LSS (XMM-Large Scale Structure; 4.7 deg$^2$). For an overview of LSST and the DDFs, see, e.g., \cite{Ivesic+19_lsst} and \cite{Brandt+18_ddfs}. 

The \mbox{X-ray} point-source catalogs for XMM-SERVS are presented in \citet[XMM-LSS]{Chen+18} and \citet[W-CDF-S and ELAIS-S1]{Ni+21_xmmservs}. They contain 11925 \mbox{X-ray} sources in total and reach a limiting flux in the \mbox{0.5--10} keV band of $\approx10^{-14}$ erg cm$^{-2}$ s$^{-1}$. Additionally, 89\%, 87\%, and 93\% of the \mbox{X-ray} sources in the W-CDF-S, ELAIS-S1, and XMM-LSS fields possess reliable multiwavelength counterparts. A summary of the \mbox{X-ray}-to-FIR surveys/missions that have observed XMM-SERVS is provided in Table 1 of \cite{Zou+22}.

We select these 21 sources from the catalog presented by \cite{Yu+24}. \cite{Yu+24} selected 3738 DOGs in XMM-SERVS using the DOG selection criteria provided in \cite{Dey+08}, creating the largest DOG catalog to-date with high-quality multiwavelength characterization. Of these, 174 (4.6\%) are detected in \mbox{X-ray}s. Their DOGs are generally not high-$\lambda_\mathrm{Edd}$ DOGs, making them different from J1324+4501 in nature. We select DOGs from this catalog by selecting those with sufficient ($>200$) net counts, for all XMM-Newton cameras combined, for good \mbox{X-ray} spectral analysis and either spectroscopic redshifts (spec-$z$s) or photometric redshifts (photo-$z$s) with $Q_z^\mathrm{good} < 1$, where $Q_z^\mathrm{good}$ is a modified version of the $Q_z$ photo-$z$ quality indicator defined in Equation 8 of \cite{Brammer+08}. In brief, \cite{Yu+24} developed this as a photo-$z$ quality indicator to be more indicative of the photo-$z$ quality for sources with extreme colors similar to DOGs. Of our 21 sources, 2 have spec-$z$s and the remaining 19 have photo-$z$s. The photo-$z$s are public and are taken from \cite{Chen+18} for XMM-LSS and \cite{Zou+21} for W-CDF-S and ELAIS-S1. The photo-$z$s have been derived using \texttt{EAZY} (\citealt{Brammer+08}), and the SED of each source contain 26 photometric bands (15 with SNR $>$ 5) on average.

The redshift range for our sample of DOGs is $z \sim 0.98-3.0$. Five of the DOGs in our sample are also analyzed in \cite{Kayal+24} with their work covering the \mbox{X-ray} properties of some DOGs in the XMM-LSS field. Further, three (WCDFS1049, WCDFS2030, XMM00267) of the 21 sources are classified as radio AGNs via the criteria in either \cite{Zhu+23} or Zhang et al. (submitted). 

We obtain the raw observation files for our supplementary \mbox{X-ray} bright sample from the XMM-Newton Science Archive\footnote{\textsuperscript{2}\url{https://nxsa.esac.esa.int/nxsa-web/\#home}}.\textsuperscript{2} We download all available archival observations for each source, and we reduce the observations and extract the corresponding spectra following the same data reduction processes as in Section \ref{sec:j1324_red}. 

\section{X-ray Spectral Modeling} \label{sec:modeling}
To fit our sources' \mbox{X-ray} spectra, we use \texttt{XSPEC} (\citealt{Arnaud+96}) models within \texttt{sherpa v4.16.0} (\citealt{Doe+07}). We use the $W$-stat statistic within \texttt{sherpa}. We also filter out energy ranges that overlap with XMM-Newton instrumental background lines (i.e., Al K$\alpha$ at $1.45-1.54$ keV; Cu at $7.2-8.2$ keV)\footnote{\textsuperscript{3}\url{https://xmm-tools.cosmos.esa.int/external/xmm_user_support/documentation/uhb/epicintbkgd.html}}.\textsuperscript{3} We outline the models tested for J1324+4501 in Section \ref{sec: xray_j1324}, and we outline the model used for our supplementary sample in Section \ref{sec: xray_supp}. For completeness, we analyze the corresponding Chandra spectrum in our XMM-Newton spectral analysis for J1324+4501; the Chandra spectrum has limited counts but very low background. We verify that this inclusion does not significantly alter our results.

\subsection{J1324+4501} \label{sec: xray_j1324}
We begin our \mbox{X-ray} spectral fitting process for J1324+4501 with a simple power law absorbed by the Galactic absorption, expressed as \texttt{phabs*zpowerlw} where \texttt{phabs} represents the Milky Way's $N_\mathrm{H}$ and \texttt{zpowerlw} represents a redshifted power-law spectrum. We will refer to this model as the $Pow$ model. The Galactic $N_\mathrm{H}$ is fixed to the value obtained from NASA's HEASARC $N_\mathrm{H}$ calculator (\citealt{heasarc}), which yields a value of $2.09 \times 10^{20}$ cm$^{-2}$. The effective power-law photon index ($\Gamma_\mathrm{eff}$) and the power-law normalization are left as free parameters in our fit. This fit \mbox{($W$-stat/d.o.f. = 454/435)} returns a hard effective photon index of $\Gamma_\mathrm{eff} = 0.03_{-0.07}^{+0.12}$, indicating that our source is heavily obscured. 

To measure the source's intrinsic $N_\mathrm{H}$, we utilize an absorbed power law, expressed as \texttt{phabs*(zphabs*cabs*zpowerlw + constant*zpowerlw)}. We will refer to this model as the $AbsPow$ model. The first term contains the transmitted component, which is represented by a redshifted power law (\texttt{zpowerlw}) with \texttt{zphabs} accounting for the source's intrinsic absorption and \texttt{cabs} modeling the Compton-scattering losses along the line-of-sight. The second term describes a soft scattered component with \texttt{constant} describing the scattered fraction ($f_\mathrm{sc}$). The $N_\mathrm{H}$ values of \texttt{zphabs} and \texttt{cabs} are linked, and the two \texttt{zpowerlw} components are set to be the same. We also fix the power-law photon index ($\Gamma$) to 1.9, and we do not observe any material change in our results if we fix it to other reasonable values, e.g., 1.8 or 2. In this fit \mbox{($W$-stat/d.o.f. = 424/434)}, we obtain \mbox{$\log (N_\mathrm{H}/\mathrm{cm}^{-2}) = 23.44^{+0.06}_{-0.07}$}, a high \mbox{X-ray} luminosity $\log (L_\mathrm{X}$/erg s$^{-1}) = 44.78^{+0.05}_{-0.08}$. Further, we obtain an $f_\mathrm{sc} = 1.7\%$ consistent with those of Compton-thick AGNs (e.g., \citealt{Lanzuisi+15}; \citealt{Li+19}). It is worth noting that the errors on $N_\mathrm{H}$ and $L_\mathrm{X}$ in this fit are likely artificially lowered by the fixing of $\Gamma$.

\begin{table*}
    \centering
    \caption{Summary of the \mbox{X-ray} spectral fitting results for J1324+4501.}
    \begin{threeparttable}
    \begin{tabular}{ccccc}
    \hline
    \hline
       Model & $\Gamma$ & $\log (L_\mathrm{X} / \mathrm{erg} \ \mathrm{s}^{-1})$ & $\log (N_\mathrm{H} / \mathrm{cm}^{-2}$) & $W$-stat/d.o.f. \\
       \hline
       $Pow$ & $0.03_{-0.07}^{+0.12}$ & $44.11^{+0.05}_{-0.04}$ & -- & 454/435 \\
       $AbsPow$ & $1.84^{+0.41}_{-0.44}$ & $44.71^{+0.08}_{-0.12}$ & $23.43^{+0.09}_{-0.13}$ & 424/433 \\
       $Torus$ & $1.75^{+0.29}_{-0.30}$ & $44.68^{+0.07}_{-0.06}$ & $23.41^{+0.06}_{-0.06}$ & 423/431 \\
       $Refl$ & $0.03_{-0.04}^{+0.10}$ & $44.39^{+0.10}_{-0.09}$ & -- & 455/434 \\
    \hline
    \end{tabular}
    \begin{tablenotes}
        \small
            \item $Notes$: The $L_\mathrm{X}$ values of the $Pow$ and $Refl$ models are observed ones with no absorption corrections, while the $L_\mathrm{X}$ values of the $AbsPow$ and $Torus$ modes are intrinsic (i.e., absorption-corrected). 
        \end{tablenotes}
    \end{threeparttable}
    \label{tab:j1324_fits}
\end{table*}

\begin{figure}
    \centering
    \includegraphics[scale=0.4]{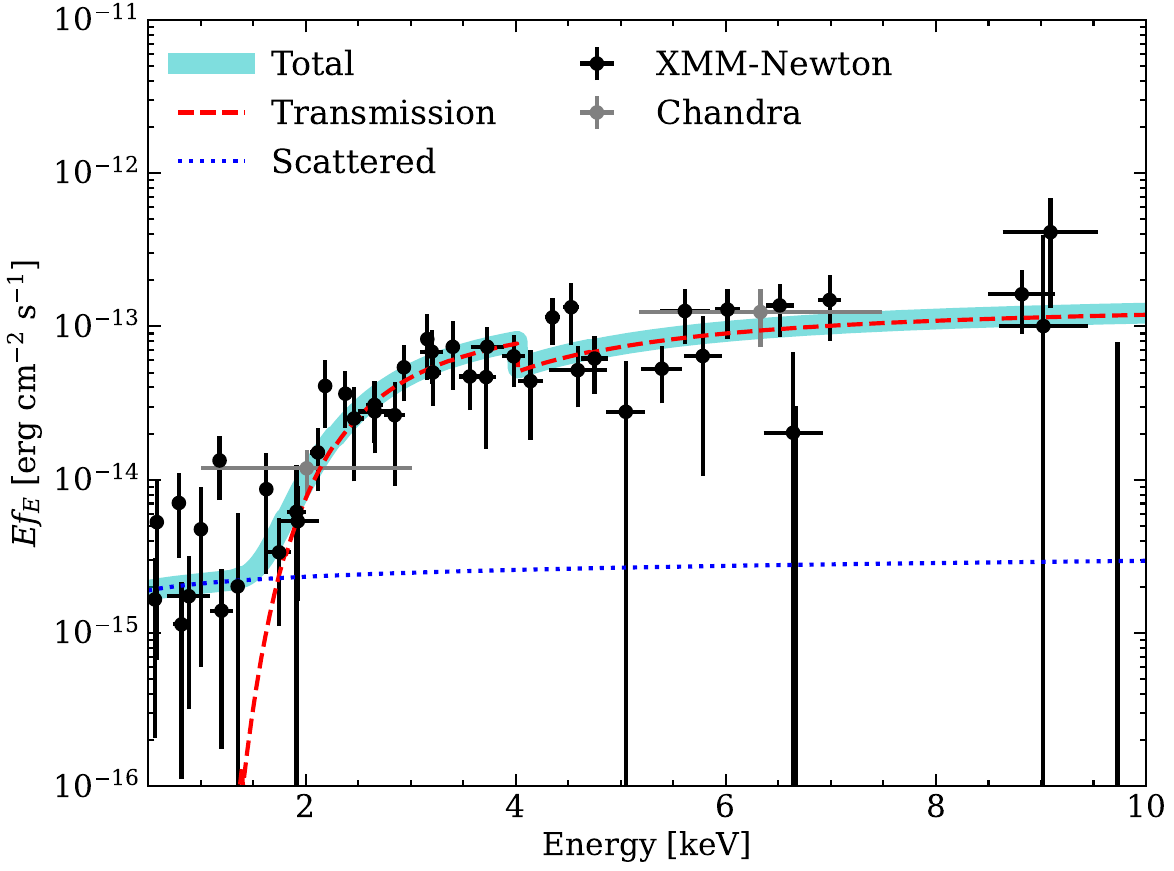}
    \caption{The unfolded joint observed-frame XMM-Newton/Chandra spectrum of J1324+4501. The best-fit $AbsPow$ model is given by the cyan line, while the individual transmission and scattered components are shown by the red-dashed and blue-dotted lines, respectively. This spectrum is rebinned for visualization purposes, and this merged spectrum is presented only for illustration purposes. Our scientific analyses are based on jointly fitting the individual spectra.}
    \label{fig:abspow_spec}
\end{figure}

Allowing $\Gamma$ to be a free parameter, we obtain $\Gamma = 1.84^{+0.41}_{-0.44}$ with \mbox{$\log (N_\mathrm{H}/\mathrm{cm}^{-2}) = 23.43^{+0.09}_{-0.13}$, $\log (L_\mathrm{X}$/erg s$^{-1}) = 44.71^{+0.08}_{-0.12}$}, and $f_\mathrm{sc} = 1.9\%$. These measurements are largely consistent with those when $\Gamma$ is fixed, and we find that allowing $\Gamma$ to vary provides a similar fit quality ($W$-stat/d.o.f. = 424/433). Because we are able to constrain $\Gamma$, we adopt these $L_\mathrm{X}$ and $N_\mathrm{H}$ values as our final measurements. We plot the ``unfolded" \mbox{X-ray} spectrum for J1324+4501 fitted with this model in Figure \ref{fig:abspow_spec}. We do not find any evidence for a neutral Fe K$\alpha$ line from the residuals of our fits. 

Third, we test a more physically motivated model in order to consider reprocessed \mbox{X-ray} emission from the circumnuclear material around the AGN (i.e., the torus; \citealt{Netzer+15}). This model is expressed as \texttt{phabs*(borus + zphabs*cabs*cutoffpl + constant*cutoffpl)} where \texttt{borus} is the reprocessed torus emission model from \cite{Balokovic+18} and \texttt{cutoffpl} is a power law with a high-energy exponential drop off. We will refer to this model as the $Torus$ model. The $N_\mathrm{H}$ values of \texttt{borus}, \texttt{zphabs}, and \texttt{cabs} are all set to be the same, and the \texttt{cutoffpl} parameters are linked to those of \texttt{borus}. Our fit returns a covering factor ($C_{tor}$) of $C_{tor} = \cos{\theta_{tor}} = 0.90^{+0.09}_{-0.10}$. We obtain $N_\mathrm{H}$, $L_\mathrm{X}$, and $f_\mathrm{sc}$ values consistent with those of our $AbsPow$ model in this fit ($W$-stat/d.o.f. =  423/431). This is due to the fact that, at $\log (N_\mathrm{H}/\mathrm{cm}^{-2}) \lesssim 23.5$, the reflected component of our spectrum is weak and the $Torus$ model then closely resembles an absorbed power-law. 

Because the \texttt{borus} model can account for the average torus $N_\mathrm{H}$ differing from the line-of-sight $N_\mathrm{H}$, we also perform a fit where we do not link the \texttt{zphabs} $N_\mathrm{H}$ (line-of-sight) and the \texttt{borus} $N_\mathrm{H}$ (average). In doing so, we obtain results consistent with those when they are linked. Because the $AbsPow$ model is simpler, we opt to use the measurements from this model as our final results for this source.

Finally, we test a reflection-dominated model ($Refl$) to assess a pure reflection case. Given the results for our $Torus$ model, we expect that the reflection-dominated model should not provide a good fit relative to our other model results. This final model is expressed as \texttt{phabs*pexrav}, where \texttt{pexrav} is the Compton-reflection model from \cite{Magdziarz+95}. For this model, we assume solar abundances and leave the inclination angle of the reflecting medium free to vary. As expected, we do not obtain a better fit than the $AbsPow$ or $Torus$ models considering this scenario ($W$-stat/d.o.f = 455/434), so we conclude that the \mbox{X-ray} spectrum of our DOG is not reflection-dominated. We provide a summary of our fitting for J1324+4501 in Table \ref{tab:j1324_fits}.

With no evidence in our fits for any Fe lines, we derive an upper-limit for the rest-frame equivalent width ($EW$) of a narrow neutral Fe K$\alpha$ line using a model expressed as \mbox{\texttt{phabs*(zphabs*cabs*zpl + zgauss + fsc*zpl)}}, where \texttt{zgauss} models a redshifted Gaussian line profile. We fix the line energy to 6.4 keV and the line width to 1 eV. We derive an upper-limit to the rest-frame $EW$ at 6.4 keV to be 0.50 keV. This rest-frame $EW$ upper-limit is smaller than rest-frame $EW$s previously observed for Hot DOGs, which are $\sim$ $1-2$ keV (e.g., \citealt{Piconcelli+15}; \citealt{Vito+18}). 

We check the reliability of our fits by creating 1000 mock spectra using our best-fit parameters and fixing them to derive their best-fit $W$-stat/d.o.f. values. To assess the quality of the fit, we compare the median $W$-stat/d.o.f. of the simulated fits to the real fit. If our fit is of good quality, the $W$-stat/d.o.f. of our fit should be close to the median $W$-stat/d.o.f. of the simulations. We indeed find that our simulations yield a median $W$-stat/d.o.f. of 0.94 with $1\sigma$ of the distribution lying between $0.86-1.04$, while the real $AbsPow$ fit yields a $W$-stat/d.o.f. = 0.98. These values are close enough to indicate that we are able to fit the data acceptably. 

Lastly, we briefly compare the results of our \mbox{X-ray} analysis to those from \cite{Zou+20}, where they fit the \mbox{X-ray} spectrum of J1324+4501 and measured a higher $\log (L_\mathrm{X} / \ \mathrm{erg \ s}^{-1}) = 45.2^{+0.2}_{-0.2}$ and $\log (N_\mathrm{H} / \mathrm{cm}^{-2}) = 23.73^{+0.14}_{-0.20}$. Their measurements are higher than ours by $\sim0.5$ dex for $L_\mathrm{X}$ and 0.3 dex for $N_\mathrm{H}$. However, the statistical uncertainties are large, especially for the Chandra observation in 2020 with only 15 counts, and thus our measurements are consistent with theirs within $2\sigma$, and any possible \mbox{X-ray} variability cannot be confirmed. 

\subsection{Supplementary Comparison Sample} \label{sec: xray_supp}
We fit our supplementary sample with the best-fit model for J1324+4501 (the $AbsPow$ model); this model is flexible and can accomodate a wide range of absorpton levels. We allow for the photon index to be free if the model is able to reasonably constrain it without too large errors (i.e., $>0.5$) or too low an index (i.e., $\Gamma < 1.7$). 

As we did for J1324+4501, we also perform 1000 simulations for each source. In doing so, we find that the median of the simulated $W$-stat/d.o.f.'s are generally consistent with the $W$-stat/d.o.f.'s from the real fits. 

\begin{table*}
    \centering
    \begin{threeparttable}
        \renewcommand{\TPTminimum}{\linewidth}
        \caption{The basic observation information and \mbox{X-ray} spectral properties for our supplementary sample.}
        \begin{tabular}{ccccccccc} 
        \hline 
        \hline 
            XID & Redshift & FB Counts & SB Counts & HB Counts & $\Gamma$ & $\log (L_\mathrm{X} / \mathrm{erg} \ \mathrm{s}^{-1})$ & $\log (N_\mathrm{H} / \mathrm{cm}^{-2}$) & $W$-stat/d.o.f. \\
            (1) & (2) & (3) & (4) & (5) & (6) & (7) & (8) & (9) \\
        \hline 
            WCDFS0192 & 1.239$_{-0.085}^{+0.091}$ & 237 & 140 & 97 & $2.53^{+0.43}_{-0.43}$ & $44.33^{+0.47}_{-0.50}$ & $23.52^{+0.20}_{-0.34}$ & 518/562 \\
            WCDFS0808 & 1.705$_{-0.066}^{+0.064}$ & 241 & 201 & 40 & 1.9$^f$ & $43.93^{+0.19}_{-0.35}$ & $22.58^{+0.31}_{-1.36}$ & 467/490 \\
            WCDFS0950 & 1.787$_{-0.192}^{+0.139}$ & 125 & 89 & 36 & $1.85^{+0.27}_{-0.28}$ & $43.42^{+0.52}_{-0.59}$ & $20.0^l$ & 348/308 \\
            WCDFS1049 & 1.732$_{-0.039}^{+0.076}$ & 366 & 223 & 143 & 1.9$^f$ & $44.71^{+0.05}_{-0.05}$ & $22.67^{+0.13}_{-0.17}$ & 449/472 \\
            WCDFS1644 & 1.787$_{-0.105}^{+0.217}$ & 149 & 110 & 39 & 1.9$^f$ & $44.11^{+0.07}_{-0.10}$ & $22.46^{+0.14}_{-0.19}$ & 599/521 \\
            WCDFS2030 & 1.603$^s$ & 4984 & 3857 & 1127 & 2.23$^{+0.06}_{-0.06}$ & $44.22^{+0.08}_{-0.08}$ & $22.27^{+0.08}_{-0.08}$ & 3677/3986 \\
            WCDFS2561 & 1.900$_{-0.093}^{+0.094}$ & 300 & 142 & 158 & 1.9$^f$ & $44.73^{+0.04}_{-0.05}$ & $22.97^{+0.06}_{-0.06}$ & 549/604 \\
            WCDFS2775 & 1.705$_{-0.104}^{+0.053}$ & 201 & 123 & 78 & 1.9$^f$ & $44.33^{+0.06}_{-0.08}$ & $22.45^{+0.13}_{-0.16}$ & 378/437 \\
            WCDFS2862 & 2.048$_{-0.175}^{+0.042}$ & 608 & 461 & 147 & $1.88^{+0.14}_{-0.14}$ & $44.78^{+0.22}_{-0.28}$ & $22.03^{+0.17}_{-0.24}$ & 688/779 \\
            ES1272 & 1.424$_{-0.103}^{+0.060}$ & 184 & 141 & 43 & $2.54^{+0.47}_{-0.47}$ & $44.16^{+0.55}_{-0.61}$ & $22.44^{+0.17}_{-0.25}$ & 372/333 \\
            ES1312 & 1.625$_{-0.118}^{+0.198}$ & 101 & 70 & 31 & 1.9$^f$ & $44.03^{+0.21}_{-0.39}$ & $23.17^{+0.25}_{-0.54}$ & 351/323 \\
            ES1783 & 2.345$_{-0.076}^{+0.112}$ & 164 & 125 & 39 & 1.9$^f$ & $44.62^{+0.11}_{-0.15}$ & $22.84^{+0.20}_{-0.32}$ & 365/346 \\
            XMM00131 & 1.732$_{-0.099}^{+0.284}$ & 359 & 260 & 99 & $1.72^{+0.21}_{-0.22}$ & $44.07^{+0.21}_{-0.25}$ & $22.21^{+0.29}_{-0.86}$ & 460/502 \\
            XMM00267 & 2.948$_{-1.001}^{+0.080}$ & 823 &  593 & 230 & 1.9$^f$ & $45.34^{+0.07}_{-0.08}$ & $22.69^{+0.16}_{-0.27}$ & 647/751 \\
            XMM00497 & 0.986$^s$ & 656 & 146 & 510 & 1.9$^f$ & $44.89^{+0.03}_{-0.03}$ & $23.17^{+0.03}_{-0.03}$ & 724/710 \\
            XMM00860 & 1.815$_{-0.345}^{+0.116}$ & 117 & 75 & 42 & 1.9$^f$ & $44.15^{+0.15}_{-0.24}$ & $22.96^{+0.25}_{-0.54}$ & 315/297 \\
            XMM01198 & 1.787$_{-0.218}^{+0.059}$ & 143 & 104 & 39 & $2.06^{+0.34}_{-0.34}$ & $44.53^{+0.05}_{-0.07}$ & $22.20^{+0.15}_{-0.21}$ & 234/243 \\
            XMM03243 & 1.678$_{-0.062}^{+0.141}$ & 652 & 482 & 170 & $1.88^{+0.14}_{-0.14}$ & $44.51^{+0.18}_{-0.23}$ & $21.98^{+0.12}_{-0.16}$ & 622/707 \\
            XMM04114 & 1.815$_{-0.197}^{+0.055}$ & 118 & 49 & 69 & 1.9$^f$ & $44.37^{+0.08}_{-0.12}$ & $23.03^{+0.12}_{-0.15}$ & 391/387 \\
            XMM04404 & 1.652$_{-0.074}^{+0.075}$ & 175 & 149 & 26 & $2.40^{+0.22}_{-0.22}$ & $44.03^{+0.35}_{-0.32}$ & $20.0^l$ & 208/281 \\
            XMM04744 & 2.538$_{-0.391}^{+0.223}$ & 173 & 111 & 62 & 1.9$^f$ & $44.65^{+0.10}_{-0.15}$ & $22.90^{+0.24}_{-0.51}$ & 347/331 \\
        \hline 
        \end{tabular}
    \begin{tablenotes}
        \small
            \item $Notes$: (1) The XID of the object from \cite{Chen+18} or \cite{Ni+21_xmmservs}. (2) The best redshift and the corresponding $1\sigma$ uncertainty of the source. $^s$Denotes sources with spec-$z$s. All other redshifts are high-quality photo-$z$s. (3) The total, aperture-uncorrected, $0.5-10$ keV source counts. (4) The total, aperture-uncorrected, $0.5-2$ keV source counts. (5) The total, aperture-uncorrected, $2-10$ keV source counts. (6) The best-fit $\Gamma$ returned from our spectral fitting process. $^f$Denotes that $\Gamma$ was fixed for this source. (7) The $L_\mathrm{X}$ returned from our spectral fitting process. (8) The $N_\mathrm{H}$ values returned from our spectral fitting process. $^l$Indicates lower-limits for sources with $N_\mathrm{H}$ values that could not be constrained well with the data. (9) The $W$-stat/d.o.f. for the fit.
        \end{tablenotes}
    \label{tab:obs_info_ext_dogs}
    \end{threeparttable}
\end{table*}

We also check for the existence of Fe K$\alpha$ lines, but none of our sources has apparent emission lines. Finally, we perform a brief comparison with the $N_\mathrm{H}$ and $L_\mathrm{X}$ values of \cite{Kayal+24}. We find that our measurements for both properties are largely consistent with theirs, despite their use of a different spectral model. There are two outliers for $L_\mathrm{X}$, XMM00131 and XMM04404. The outlier for XMM00131 is likely caused by our decision to let $\Gamma$ vary for this source, while \cite{Kayal+24} fix $\Gamma$ to 2.0. For the outlier for XMM04404, its spectrum is soft, and our best-fit $N_\mathrm{H}$ value reaches the lower limit of $N_\mathrm{H}=10 ^{20}~\mathrm{cm^{-2}}$. With a steep $\Gamma$ (see Table \ref{tab:obs_info_ext_dogs}), this source may have a strong soft-excess component, leading to our $AbsPow$ model being unable to constrain $N_\mathrm{H}$. We have double-checked our procedures and made sure that our results should be reliable.

Finally, we provide a table summarizing the results for each source in Table \ref{tab:obs_info_ext_dogs}. Note that one source, WCDFS2030, has nearly 5000 aperture-uncorrected source counts across the 34 archival observations used in our work. This is due to this source lying in the original CDF-S proper where XMM-CDF-S observations were very deep (e.g., \citealt{Ranalli+13}; Figure 1 in \citealt{Ni+21_xmmservs}). 
    
\section{Multiwavelength SED of J1324+4501} \label{sec:hosts}
To further characterize J1324+4501, we fit its SED using the available photometry from the literature in order to obtain measurements for, e.g., the host-galaxy stellar mass ($M_\star$) and star-formation rate (SFR). While this source's SED is modeled in \cite{Zou+20}, we revisit and revise the SED fitting results for two reasons. First, the \mbox{X-ray} flux used in the fitting in \cite{Zou+20} is based upon the limited counts from Chandra. We refit the SED of J1324+4501 using our new \mbox{X-ray} flux value to best constrain the AGN component of our DOG's SED. Second, \cite{Zou+20} used a delayed star formation history (SFH) to measure the source's SFR and obtained a very high SFR ($\sim 140$ $M_\odot$ yr$^{-1}$). This indicates that the AGN is residing in a galaxy producing stars at an extremely rapid rate (i.e., a ``starburst" galaxy). For this reason, we opt to use a truncated delayed SFH to better model this starburst activity.   

J1324+4501 has photometry in the SDSS DR12 $ugriz$, Pan-STARRS1 $griz$, WISE 3.4, 4.6, 12, and 22 $\mu$m, and AKARI 9, 18, 65, 90, 140, and 160 $\mu$m bands. However, J1324+4501 is not detected by AKARI (\citealt{Toba+16}), and thus we adopt the 5$\sigma$ photometric upper limits: 0.05, 0.12, 2.4, 0.55, 1.4, and 6.2 Jy in each band, respectively (\citealt{Kawada+07}; \citealt{Ishihara+10}). We further impose a 5\% error floor when fitting the photometry, and we correct for Galactic extinction using the extinction coefficients provided in \cite{Zhang_R+23}. The UV-to-IR photometric measurements used in our modeling are presented in Table \ref{tab:photometry}.

\begin{table}
    \centering
    \begin{threeparttable}
        \renewcommand{\TPTminimum}{\linewidth}
        \caption{Multiwavelength photometry of J1324+4501.}
        \begin{tabular}{cc|cc} 
            \hline\hline
            ~\rule{0.2in}{0in} & Band & Magnitude (AB) & ~\rule{0.2in}{0in} \\
            \hline
            ~\rule{0.2in}{0in} & SDSS $u$ & $22.87 \pm 0.58$ \\
            ~\rule{0.2in}{0in} & SDSS $g$ & $21.97 \pm 0.12$ \\
            ~\rule{0.2in}{0in} & SDSS $r$ & $21.46 \pm 0.12$ \\
            ~\rule{0.2in}{0in} & SDSS $i$ & $20.95 \pm 0.13$ \\
            ~\rule{0.2in}{0in} & SDSS $z$ & $19.83 \pm 0.14$ \\ 
            ~\rule{0.2in}{0in} & PS1 $g$ & $21.97 \pm 0.06$ \\
            ~\rule{0.2in}{0in} & PS1 $r$ & $21.22 \pm 0.05$ \\
            ~\rule{0.2in}{0in} & PS1 $i$ & $20.86 \pm 0.07$ \\
            ~\rule{0.2in}{0in} & PS1 $z$ & $19.81 \pm 0.11$ \\
            ~\rule{0.2in}{0in} & WISE W1 & $16.95 \pm 0.11$ \\
            ~\rule{0.2in}{0in} & WISE W2 & $16.02 \pm 0.11$ \\
            ~\rule{0.2in}{0in} & WISE W3 & $14.35 \pm 0.11$ \\
            ~\rule{0.2in}{0in} & WISE W4 & $13.14 \pm 0.12$ \\
            ~\rule{0.2in}{0in} & AKARI S9W & $12.15$$^u$ \\
            ~\rule{0.2in}{0in} & AKARI L18W & $11.20$$^u$ \\
            ~\rule{0.2in}{0in} & AKARI N60 & $7.95$$^u$ \\
            ~\rule{0.2in}{0in} & AKARI Wide-S & $9.54$$^u$ \\
            ~\rule{0.2in}{0in} & AKARI Wide-L & $8.53$$^u$ \\
            ~\rule{0.2in}{0in} & AKARI N160 & $6.90$$^u$ \\
            \hline 
        \end{tabular}
        \begin{tablenotes}
        \small
            \item $Notes$: The photometry has been corrected for Galactic extinction. Magnitudes and their uncertainties are quoted in the AB system (\citealt{Oke+83}). $^u$Denotes values that are upper-limits.
        \end{tablenotes}
        \label{tab:photometry}
    \end{threeparttable} 
\end{table}

\begin{figure*}
    \centering
    \includegraphics[scale=0.4]{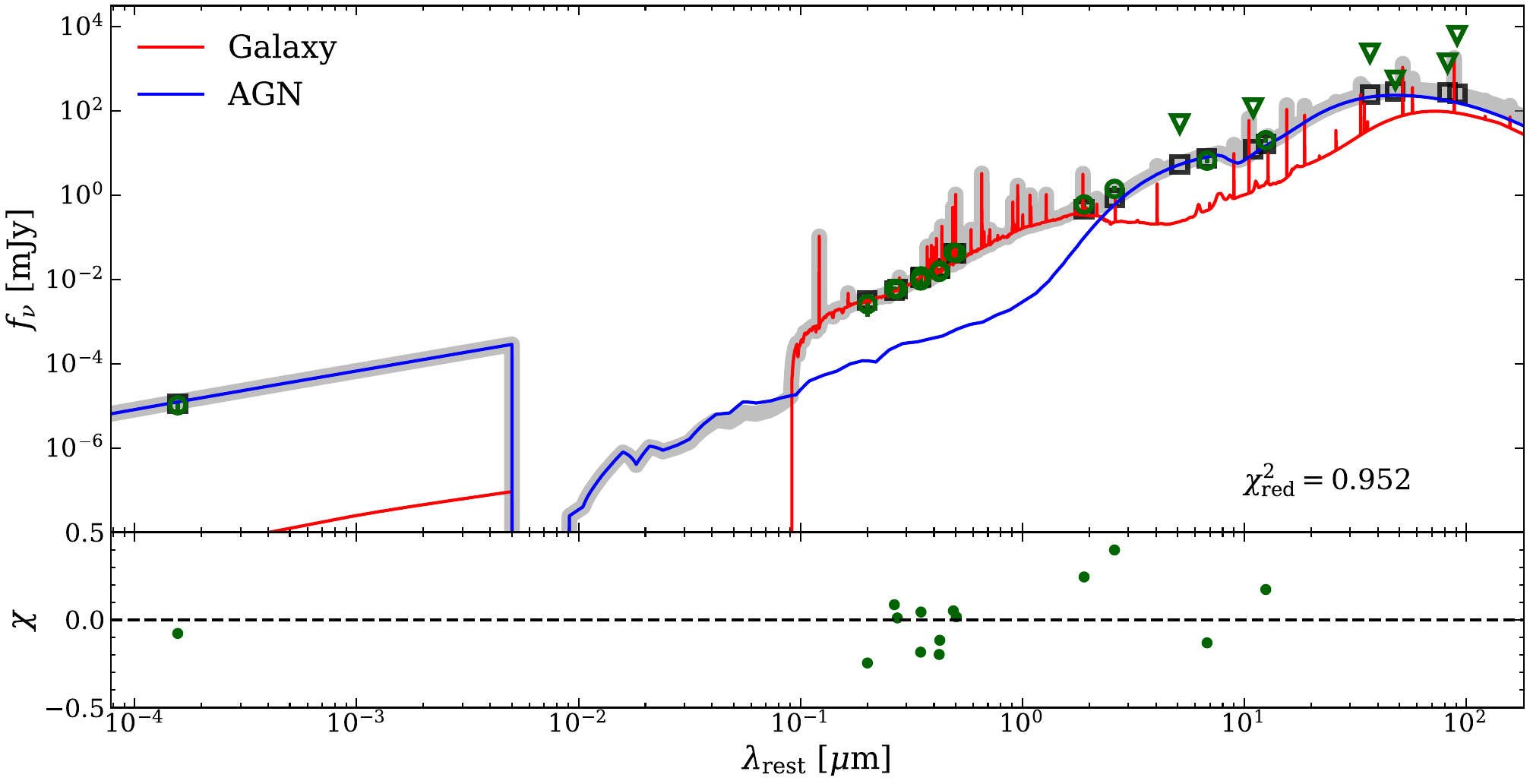}
    \caption{The best-fit SED for J1324+4501. The observed photometry is shown by the green points, the AKARI upper-limits are shown by the green downward-facing triangles, and the model spectrum (photometry) is shown by the grey line (black squares). The SED is decomposed into the galaxy components (red) and the AGN components (blue), and the reduced chi-squared of the fit is shown in the lower right. The bottom panel shows the $\chi$ residuals of the fit for points that are not upper-limits.}
    \label{fig:sed}
\end{figure*}

We utilize \texttt{CIGALE v2022.1} (\citealt{Boquien+19}; \citealt{Yang+20_cigale, Yang+22_cigale}) to model the SED, where the AGN component and its X-ray emission can be appropriately treated. We adopt a truncated delayed SFH (\citealt{Ciesla+16}), modeled by 
\begin{equation*} \label{eq:model3}
\mathrm{SFR}(t) \propto
\begin{cases}
    t \times e^{-t / \tau}, &  t \le t_\mathrm{trunc}\\
    r_\mathrm{SFR} \times \mathrm{SFR}(t_\mathrm{trunc}), & t > t_\mathrm{trunc}
\end{cases}
\end{equation*} \\
where the formula at $t \le t_\mathrm{trunc}$ is the normal delayed SFH with an $e$-folding time of $\tau$, and the SFR is assumed to instantaneously change by a factor of $r_\mathrm{SFR}$ at $t_\mathrm{trunc}$ and remain constant until the current age. Stellar templates are from \cite{Bruzual+03}, assuming a \cite{Chabrier+03} initial mass function. The host-galaxy dust attenuation is assumed to follow \cite{Calzetti+00}, and the IR dust emission follows the models in \cite{Draine+14}, updated from \cite{Draine+07} based upon detailed observations of M31. The UV-to-IR AGN module is based on the SKIRTOR model (\citealt{Stalevski+12, Stalevski+16}) with polar dust following the extinction law in the Small Magellanic Cloud (\citealt{Prevot+84}). We allow for torus inclination angles of $30\degree$ (Type 1) and $70\degree$ (Type 2), and we let $\alpha_\mathrm{ox}$ vary between $-1.9$ and $-1.1$ in \mbox{increments of 0.2}.\footnote{\textsuperscript{4}$\alpha_\mathrm{ox}= -0.3838 \log (L_\mathrm{2500 \ \AA} / L_\mathrm{2 \ keV})$ is the SED slope between the UV and X-ray and connects the X-ray photometry to the AGN disk emission in \texttt{CIGALE}.}\textsuperscript{4} The disk spectral shape is modified from \cite{Schartmann+05}. We plot the best-fit SED in Figure \ref{fig:sed}.

The SEDs of the 21 DOGs in our supplementary comparison sample are fitted in \cite{Zou+22} using \texttt{CIGALE v2022.0}. We utilize their host-galaxy measurements for the sources in this sample.

Through our SED fitting for J1324+4501, we measure an extremely high instantaneous SFR of \mbox{$\log (\mathrm{SFR} / M_\odot \ \mathrm{yr}^{-1}) = 2.68 \pm 0.43$}. We also find that the host galaxy of J1324+4501 is very massive, with a stellar mass of $\log (M_\star / M_\odot) = 11.29 \pm 0.51$. With the black-hole mass measurement from \cite{Zou+20} of $\log (M_\mathrm{BH} / M_\odot) = 8.27 \pm 0.40$, measured through its broad \mbox{Mg \rom{2}} emission line, we find that the central SMBH is \mbox{$\sim 0.1\%$} of the total mass. This is consistent with the local relation from \cite{Reines+15}. We further measure the fractional flux contribution by the AGN in the IR to be $f_\mathrm{AGN} = (59 \pm 13)\%$ and the best-fit SED to favor a Type 2 solution.

The extreme host-galaxy and AGN nature of J1324+4501 forces us to take the above SED fitting results into careful consideration. From the best-fit SED, it is clear that the galaxy component dominates below rest-frame $\approx 1$ $\mu$m via stellar emission, while the AGN component dominates above rest-frame $\approx 3$ $\mu$m via the AGN-heated dust. However, discerning between the AGN-heated dust and the old stellar population in the wavelength range between this ($\approx 1-3$ $\mu$m) becomes difficult. This is the wavelength range in which the old stellar emission or the AGN component could be dominant. Thus, constraining the true SFR, $M_\star$, and AGN fraction becomes difficult, because the measured $M_\star$ and AGN fraction may be dependent on the assumptions about the source's SFH (i.e., how much old stellar emission dominates).

\cite{Zou+20} performed SED fitting for J1324+4501, measuring a much lower $\log (M_\star \ / \ M_\odot) = 10.50 \pm 0.13$ and an instantaneous $\log (\mathrm{SFR} \ / \ M_\odot \ \mathrm{yr}^{-1}) = 2.15 \pm 0.11$. Their $f_\mathrm{AGN}$ measurement was much higher, however, at $f_\mathrm{AGN} = (81.7 \pm 7.6)\%$. The lower bound on our $M_\star$ measurement does not rule out the \cite{Zou+20} measurement. Indeed, the \cite{Zou+20} measurement may be viewed as a lower limit on the source's $M_\star$ since a delayed SFH assumes a younger stellar population without many old stars. On the other hand, the differences between SFR and $f_\mathrm{AGN}$ are more notable. One of the reasons for the $f_\mathrm{AGN}$ difference could be due to the \cite{Zou+20} $L_\mathrm{X}$ measurement being much higher than ours. A second reason, as discussed above, could be the choice of SFH and the subsequent stellar-population assumptions, which also may explain the difference in measured SFR.

To test what is affecting the recovered SED results the most, we perform two checks. First, we fit the SED using our UV-to-IR data but with the X-ray measurement from \cite{Zou+20} to test whether the discrepancy is due to the differences in $L_\mathrm{X}$. In this fit, we obtain $M_\star$, SFR, and $f_\mathrm{AGN}$ values close to those recovered with our X-ray measurement. Thus, the X-ray measurement is not a large factor in the discrepancy. Second, we fit the SED using the original \cite{Zou+20} data and our stellar-population assumptions. This fit returns a higher $\log (M_\star \ / \ M_\odot) = 11.56 \pm 0.34$ and $\log (\mathrm{SFR} \ / \ M_\odot \ \mathrm{yr}^{-1}) = 2.75 \pm 0.43$, and a lower $f_\mathrm{AGN} = (53.5 \pm 20)\%$. These values are still consistent with ours; thus, we conclude that the stellar-population assumptions likely play a key role in the returned $M_\star$, SFR, and $f_\mathrm{AGN}$ values returned from \texttt{CIGALE} for this source.

To further test the reliability of our recovered $M_\star$ and SFR values with \texttt{CIGALE}, we fit the optical photometry using the \texttt{Prospector}-$\alpha$ model within \texttt{Prospector} (\citealt{Leja+17}; \citealt{Johnson+21}), which includes complex dust attenuation and re-radiation, nebular emission, gas- and stellar-phase metallicity, and a six-component nonparametric SFH. We choose this model both for its flexibility and its nonparametric SFH. The nonparametric SFH is particularly important as it allows us not to rest on strong assumptions concerning the source's SFH, which should reduce biases in the recovered $M_\star$ or SFR (\citealt{Leja+19a}). We opt not to include the WISE photometry into the fitting because AGN-dominated IR colors may be mistaken as circumstellar dust around AGB stars in the fitting and affect the recovered SFH (e.g., \citealt{Leja+17}). 

\begin{figure}
    \centering
    \includegraphics[scale=0.5]{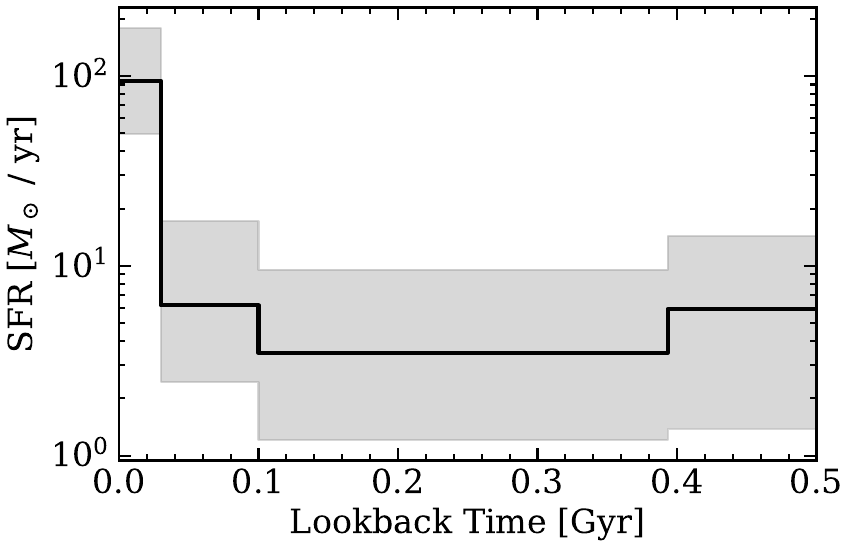}
    \caption{The recovered SFH from our \texttt{Prospector}-$\alpha$ fit. The 50th percentile of the SFH is shown in black, while the 16th and 84th percentiles of the SFH are shown in grey.}
    \label{fig:sfh}
\end{figure}

This fit returns values of \mbox{$\log (\mathrm{SFR} / M_\odot \ \mathrm{yr}^{-1}) = 1.97^{+0.28}_{-0.27}$} and \mbox{$\log (M_\star / M_\odot) = 11.04^{+0.16}_{-0.13}$}. Interestingly, the recovered SFH suggests a recent starburst (see Figure \ref{fig:sfh}), which indicates that our choice in SFH in our \texttt{CIGALE} fit was reasonable. The SFR is lower than that of our \texttt{CIGALE} fit by a factor of $\approx5$, but it possesses errors of similar proportion. Choosing to remove the WISE photometry may have lowered the recovered SFR for this fit, as we may be losing important information on re-radiated UV emission. From our \texttt{CIGALE} fit, we obtain a best-fit $E(B-V)=0.5$, suggesting that there is a large amount of UV radiation being lost to attentuation and re-radiated in the IR. This large difference in SFR may be due to the fact that \texttt{CIGALE} is able to account for both the AGN and galaxy components in the IR, while our \texttt{Prospector} fit does not take into account either the AGN component or the strong UV attenuation/re-radiation. Thus, our results are not materially weakened by this factor as our \texttt{CIGALE} fit is the best model to use for this source. 

It is worth noting that while the SFH suggests a starburst, recent works focusing on Hot DOGs with excess UV emission (BluDOGs; e.g., \citealt{Nob+22, Nob+23}) have found that the UV emission may be attributed to either a strong starburst or leaked UV emission from the central AGN that has been scattered into our line of sight (e.g., \citealt{Assef+16, Assef+20}). J1324+4501 shares some similarities with BluDOGs, so we take the SFR results with caution.

Based upon our new SED fitting and the $M_\mathrm{BH}$ measurement from \cite{Zou+20}, we are able to estimate $\lambda_\mathrm{Edd}$ for J1324+4501 via the equation 
\begin{equation*}
    \lambda_\mathrm{Edd} = \frac{L_\mathrm{bol} \ / \ (\mathrm{erg} \ \mathrm{s}^{-1})}{1.26 \times 10^{38} \ M_\mathrm{BH} \ / \  M_\odot}
\end{equation*}
where $L_\mathrm{bol}$ is estimated within \texttt{CIGALE} and errors are propagated from both $M_\mathrm{BH}$ and $L_\mathrm{bol}$. We estimate a value of $\lambda_\mathrm{Edd}=1.13_{-0.71}^{+1.34}$, placing this DOG on the boundary of the expected area for (Hot) DOGs in the $\lambda_\mathrm{Edd}-N_\mathrm{H}$ plane (e.g., \citealt{Ishibashi+18}; \citealt{Wu+18}; \citealt{Vito+18}; Figure 7 of \citealt{Zou+20}). The large error bars in our measurement of $\lambda_\mathrm{Edd}$ are dominated by the uncertainties in our black hole mass measurement (\citealt{Zou+20}). From a physical standpoint, this indicates that J1324+4501 may be in a similar evolutionary phase as Hot DOGs, where the central AGN is at the peak of its growth and has not entered the blow-out phase. 

\section{Results} \label{sec:results}
In this section, we analyze the basic physical characteristics of our DOGs in comparison to those previously reported in the literature, investigate the nature of the obscuring material in our DOGs, and study the host-galaxy star formation of J1324+4501 relative to the general DOG population. 

\subsection{The $N_\mathrm{H}-L_\mathrm{X}$ Plane}
\begin{figure}
    \centering
    \includegraphics[scale=0.5]{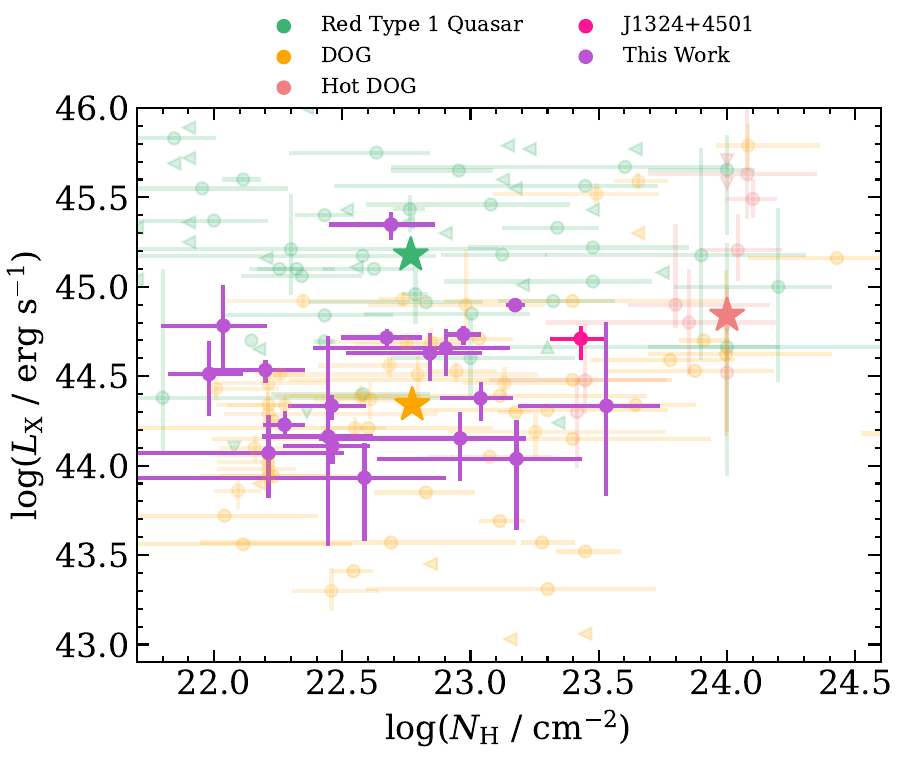}
    \caption{The $N_\mathrm{H}-L_\mathrm{X}$ plane for several types of AGNs. Our primary source, J1324+4501, is plotted in pink. Our \mbox{X-ray} bright supplementary sample is plotted in purple. Reddened Type 1 quasars, DOGs, and Hot DOGs from the literature (see text) are plotted in light green, orange, and red, respectively. The median $L_\mathrm{X}$ and $N_\mathrm{H}$ values for each population are shown by the light green, orange, and red stars.}
    \label{fig:nh_lx}
\end{figure}

We plot our sources in the $N_\mathrm{H}-L_\mathrm{X}$ plane in Figure \ref{fig:nh_lx}. For reference, we plot reddened type 1 quasars (\citealt{Urrutia+05}; \citealt{Martocchia+17}; \citealt{Mountrichas+17}; \citealt{Goulding+18}; \citealt{Lansbury+20}), DOGs (\citealt{Lanzuisi+09}; \citealt{Corral+16}; \citealt{Zou+20}; \citealt{Kayal+24}), and Hot DOGs (\citealt{Stern+14}; \citealt{Assef+16}; \citealt{Ricci+17}; \citealt{Vito+18}; \citealt{Zappacosta+18}).  

We find that J1324+4501 is one of the most obscured DOGs in our sample and lies in a region of the plane where some Hot DOGs live. This is an indicator that the physical nature of J1324+4501, as a high-$\lambda_\mathrm{Edd}$ DOG, differs from the general DOG population. It is likely closer to the peak of its SMBH accretion and obscuration and is in a similar evolutionary phase as Hot DOGs. It is less obscured than most of the Hot DOGs (with most Hot DOGs reaching the Compton-thick level), but possesses a similar $L_\mathrm{X}$ as the lower $L_\mathrm{X}$ Hot DOGs. 

On the other hand, our supplementary DOGs are largely physically similar to the general DOG population that contains AGNs. They span a wide range of column densities and \mbox{X-ray} luminosities, highlighting the heterogeneous nature of DOGs. There is one object, XMM00267, which possesses an extremely high $L_\mathrm{X}$ with a moderate $N_\mathrm{H}$, placing it in the ``Red Type 1 Quasar" area of the diagram. This DOG likely is past its obscuration peak with much of its obscuring material having been swept out already, and it is the only DOG in our supplementary sample of such nature. 

\subsection{The $L_\mathrm{6\mu\mathrm{m}}-L_\mathrm{X}$ Plane}
We utilize the $L_\mathrm{6\mu\mathrm{m}}-L_\mathrm{X}$ diagnostic plot to investigate the nature of the obscuring material in our DOGs, where $L_\mathrm{6\mu\mathrm{m}}$ is defined as $\nu L_\nu$ at rest-frame 6 $\mu$m (for the AGN component only), and $L_\mathrm{X, obs}$ is the observed $L_\mathrm{X}$ at rest-frame \mbox{$2-10$ keV} (i.e., not corrected for intrinsic absorption). We estimate $L_\mathrm{X, obs}$ for each of our sources using a simple model expressed as \texttt{phabs*zpowerlw} (i.e., the $Pow$ model for J1324+4501). 

If a source possesses heavy intrinsic absorption, the observed \mbox{X-ray} emission is expected to be lower relative to the intrinsic \mbox{X-ray} emission while the 6 $\mu$m emission should remain largely the same. We estimate $L_\mathrm{6\mu\mathrm{m}}$ for J1324+4501 through the SED-fitting results. We have WISE photometric measurements around rest-frame 6 $\mu$m (see Figure \ref{fig:sed}), so we should be able to constrain this luminosity well in the fitting process. For our supplementary sample, we utilize the $L_\mathrm{6\mu\mathrm{m}}$ measurements provided in the best-fit SEDs of \cite{Yu+24}, who obtained them from \cite{Zou+22}. 

\begin{figure}
    \centering
    \includegraphics[scale=0.5]{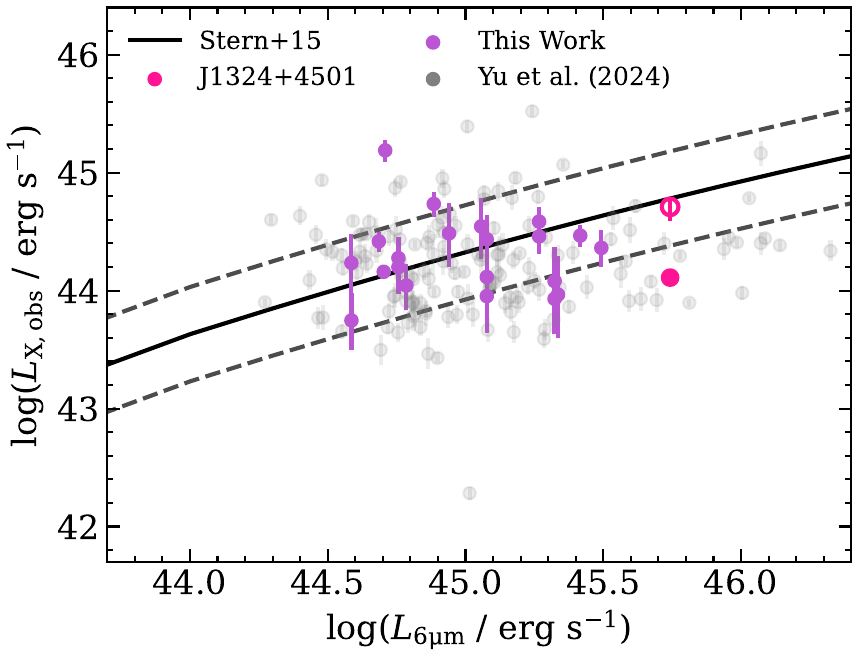}
    \caption{Our sources in the $L_\mathrm{6\mu\mathrm{m}}-L_\mathrm{X, obs}$ plane. Similar to \mbox{Figure \ref{fig:nh_lx}}, J1324+4501 is plotted in pink, and our supplementary sample is plotted in purple. The open pink circle shows the intrinsic $L_\mathrm{X}$ for J1324+4501. For reference, we also include the DOGs from \cite{Yu+24} in grey. The solid black line is the relation from \cite{Stern+15}, and the dashed black lines represent the $1\sigma$ deviation from this relation.}
    \label{fig:lx_l6}
\end{figure}

We plot our DOGs' $L_\mathrm{X, obs}$ against their AGN $L_\mathrm{6\mu\mathrm{m}}$ in Figure \ref{fig:lx_l6} in addition to the X-ray detected DOGs from \cite{Yu+24} for reference. We find that the majority of our sources are consistent (within 1$\sigma$) with the \cite{Stern+15} relation, and the majority of our supplementary sample has small absorption corrections. However, there are two notable deviations in J1324+4501 and XMM00267.

Recalling the intrinsic $L_\mathrm{X}$ measurements from \mbox{Table \ref{tab:j1324_fits}}, Figure \ref{fig:lx_l6} demonstrates the heavy \mbox{X-ray} absorption in J1324+4501. There is a significant $\sim 0.6$ dex change between its intrinsic $L_\mathrm{X}$ and $L_\mathrm{X, obs}$. J1324+4501 is also very luminous in the IR, with a high $L_\mathrm{6\mu\mathrm{m}}$ relative to the other DOGs in our sample. 

XMM00267 is once again a source to note relative to the other supplementary DOGs, having a large $L_\mathrm{X, obs}$. It has a minimal deviation from its intrinsic luminosity, with a $\sim 0.1$ dex change between the quantities. Given that this source is also detected at radio wavelengths, enhanced coronal emission and/or jet-linked emission may be responsible for its elevated $L_\mathrm{X}$ (e.g., \citealt{Worrall+87}; \citealt{Miller+11}; \citealt{Zhu+20, Zhu+21}). Therefore, with its placement in the $N_\mathrm{H}-L_\mathrm{X}$ plane in mind, we can interpret this source as a likely post-merger galaxy, past its obscuration peak, likely close to becoming an unobscured quasar. 

\subsection{Host-Galaxy Star Formation}
The SED fitting in Section \ref{sec:hosts} provides us with host-galaxy measurements for J1324+4501. To compare the host galaxy of J1324+4501 to other DOGs, we utilize the catalog from \cite{Yu+24}. \cite{Yu+24} selected $\sim3700$ DOGs in the XMM-SERVS fields and studied their basic \mbox{X-ray} and host-galaxy properties, providing the largest catalog of well-characterized DOGs to-date. The SED results utilized in \cite{Yu+24} also come from \cite{Zou+22}. 

Rather than investigate the SFRs of our DOGs directly, we analyze their SFRs relative to the star-forming main sequence (MS) predicted SFRs (SFR$_\mathrm{MS}$). To do so, we use the normalized SFR (SFR$_\mathrm{norm}$; $\frac{
\mathrm{SFR}}{\mathrm{SFR}_\mathrm{MS}}$) values from \cite{Yu+24} to calculate $\mathrm{SFR}_\mathrm{MS}$ for our supplementary sample. Several previous works have successfully utilized this value to indirectly study the relationship between SFR and other properties of the central AGN and/or the host galaxy (e.g., \citealt{Mullaney+15};  \citealt{Mountrichas+21}; \citealt{Vietri+22}; \citealt{Birchall+23}; \citealt{Cristello+24}; \citealt{Mountrichas+24b}; \citealt{Yu+24}; Zhang et al. submitted). We calculate SFR$_\mathrm{MS}$ for J1324+4501 by utilizing the SFR and $M_\star$ from our \texttt{CIGALE} fitting and the MS from \cite{Popesso+23}. We calculate the errors in SFR$_\mathrm{MS}$ using 
\begin{equation*}
    \sigma_{\log \mathrm{SFR_{MS}}} = \left|\frac{d\log \mathrm{SFR}}{d\log M_\star}\right| \times \sigma_{\log M_\star}
\end{equation*}
where $\frac{d\log \mathrm{SFR}}{d\log M_\star}$ is the local slope of the main sequence at a combination of ($z$, $M_\star$). These errors are generally small.

\begin{figure}
    \centering
    \includegraphics[scale=0.5]{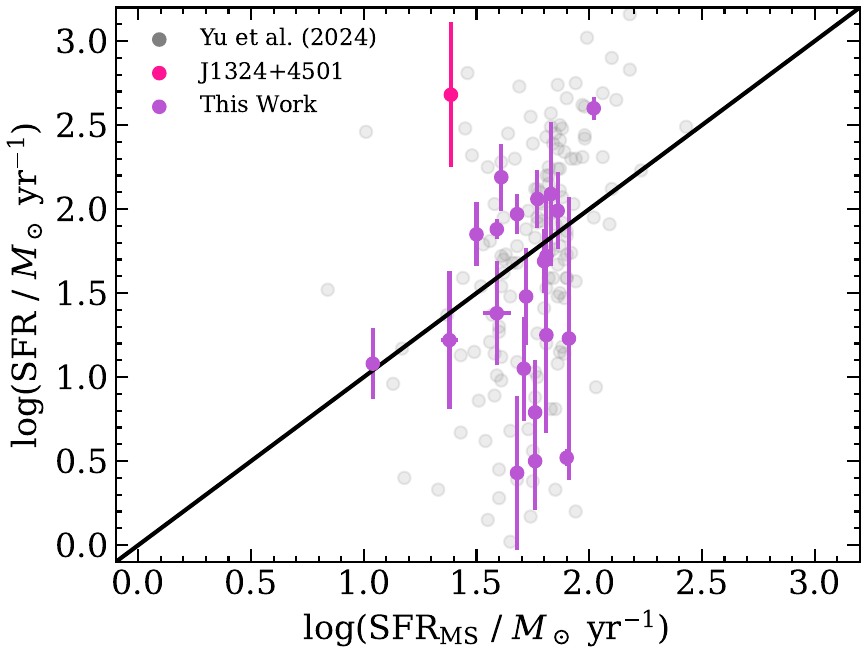}
    \caption{A comparison between SFR$_\mathrm{MS}$ and SFR for our sample. J1324+4501 is shown in pink, our supplementary sample in purple, and the \mbox{X-ray} detected DOGs from \cite{Yu+24} in grey. The black solid line represents the 1-to-1 line.}
    \label{fig:ms_comp}
\end{figure}

We plot the measured SFR against the SFR$_\mathrm{MS}$ for our supplementary DOG sample, in addition to the \mbox{X-ray} detected DOGs from \cite{Yu+24}, in Figure \ref{fig:ms_comp}. In the figure, it is clear that the DOG population possesses a wide range of SFRs across an order of magnitude of MS-predicted SFRs. We find that many of the DOGs in our supplementary sample are consistent with their MS-predicted values. However, we find that J1324+4501 is undergoing more-intense starburst activity than the general DOG population, with its actual SFR residing roughly an order of magnitude above the MS SFR. This remarkable starburst activity is consistent with other high-$\lambda_\mathrm{Edd}$ DOGs (e.g. \citealt{Zou+20}), indicating that J1324+4501 is also at the peak of galaxy growth among gas-rich mergers.

\section{Discussion and Conclusions} \label{sec:disc}
In this work, we utilized 22 \mbox{X-ray} luminous DOGs to provide the most-detailed \mbox{X-ray} spectral analysis of \mbox{X-ray} detected DOGs to-date. We find that the spectra of all our sources are well-represented by a redshifted, absorbed power-law with a soft scattering component. 

J1324+4501 is the most exceptional source with a high $\lambda_\mathrm{Edd}$, $L_\mathrm{X}$, and $N_\mathrm{H}$. We perform SED fitting for this source using the available archival photometry, finding that it is experiencing an extreme starburst unseen in many other \mbox{X-ray} detected DOGs. Due to the extreme nature of the source, it is difficult to place robust constraints on the stellar mass, SFR, or AGN fraction, since these quantities may depend on assumptions about the source's SFH. The nature of J1324+4501 indicates that it is physically different from typical DOGs, and its high-$\lambda_\mathrm{Edd}$ nature pushes it closer to the Hot DOG population. 

The 21 other \mbox{X-ray} luminous DOGs are largely physically consistent with the general DOG population that contains AGNs. There is one exception in XMM00267, with this source showing more characteristics consistent with red, Type 1 quasars rather than DOGs. Even with a selection effect that biases our sample toward high-$L_\mathrm{X}$/low-$N_\mathrm{H}$ sources, we still observe a wide range in both $L_\mathrm{X}$ and $N_\mathrm{H}$ for these sources. 

Overall, our comparison between J1324+4501 and other \mbox{X-ray} luminous DOGs can be linked to the SMBH-galaxy coevolution framework. This work demonstrates that our high-$\lambda_\mathrm{Edd}$ DOG is indeed in a similar evolutionary stage as Hot DOGs, where accretion and star formation are at their peak. At the same time, all other DOGs in our sample show a wide range in properties such as $L_\mathrm{X}$, $N_\mathrm{H}$, and SFR, suggesting that the general DOG population is more heterogeneous in nature. This comparison also suggests that $\lambda_\mathrm{Edd}$ may be a key factor in distinguishing between different kinds of DOGs (\citealt{Zou+20}). A larger sample of high- or low-$\lambda_\mathrm{Edd}$ DOGs with quality X-ray spectroscopy would yield a useful comparison between DOGs and provide insight into this idea.

Most DOGs require $\approx 50-100$ ks of observing time with XMM-Newton or Chandra to acquire sufficient counts for X-ray analysis. For high-$\lambda_\mathrm{Edd}$ DOGs in particular, which are more extreme and more obscured, even much longer exposures are necessary. However, next-generation X-ray observatories such as Athena (\citealt{Nandra+13}) and Lynx (\citealt{Gaskin+19}) will provide unprecedented opportunities to probe the X-ray properties of DOGs with their high throughput, shortening the required observing time by a factor of ten. The Athena and Lynx X-ray microcalorimeters will also provide unprecedented spectral resolving power, allowing the detection of, e.g., Fe lines and resolution of spectral features that may otherwise be blended. 

Further, JWST imaging and spectroscopy with NIRCam, NIRSpec, and MIRI would enable detailed analyses of the infrared properties of J1324+4501 and (Hot) DOGs more generally. In the case of J1324+4501, these data would enable high-quality analysis of rest-frame $\sim 0.3-16$ $\mu$m emission, providing a detailed decomposition of the galaxy and AGN contributions to the SED. In turn, it would be possible to determine more-accurate constraints on, e.g., $M_\star$ and SFR, and would provide important information on both the role of the AGN and the galaxy in the evolution of this object. 

\section*{Acknowledgements}
We thank the anonymous referee for a thorough and constructive review, which has greatly improved this article. We thank Wei Yan, Joel Leja, and Bingjie Wang for helpful discussions. NC, FZ, WNB, and ZY acknowledge support from the Penn State Eberly Endowment, NSF grants AST-2106990 and AST-2407089, and Chandra X-ray Center grant AR4-25008X. 

\bibliography{sample631}{} 

\begin{thebibliography}{}
\expandafter\ifx\csname natexlab\endcsname\relax\def\natexlab#1{#1}\fi
\providecommand{\url}[1]{\href{#1}{#1}}
\providecommand{\dodoi}[1]{doi:~\href{http://doi.org/#1}{\nolinkurl{#1}}}
\providecommand{\doeprint}[1]{\href{http://ascl.net/#1}{\nolinkurl{http://ascl.net/#1}}}
\providecommand{\doarXiv}[1]{\href{https://arxiv.org/abs/#1}{\nolinkurl{https://arxiv.org/abs/#1}}}

\bibitem[{{Alexander} \& {Hickox}(2012)}]{Alexander+12}
{Alexander}, D.~M., \& {Hickox}, R.~C. 2012, \nar, 56, 93, \dodoi{10.1016/j.newar.2011.11.003}

\bibitem[{{Arnaud}(1996)}]{Arnaud+96}
{Arnaud}, K.~A. 1996, in Astronomical Society of the Pacific Conference Series, Vol. 101, Astronomical Data Analysis Software and Systems V, ed. G.~H. {Jacoby} \& J.~{Barnes}, 17

\bibitem[{{Assef} {et~al.}(2016){Assef}, {Walton}, {Brightman}, {Stern}, {Alexander}, {Bauer}, {Blain}, {Diaz-Santos}, {Eisenhardt}, {Finkelstein}, {Hickox}, {Tsai}, \& {Wu}}]{Assef+16}
{Assef}, R.~J., {Walton}, D.~J., {Brightman}, M., {et~al.} 2016, \apj, 819, 111, \dodoi{10.3847/0004-637X/819/2/111}

\bibitem[{{Assef} {et~al.}(2020){Assef}, {Brightman}, {Walton}, {Stern}, {Bauer}, {Blain}, {D{\'\i}az-Santos}, {Eisenhardt}, {Hickox}, {Jun}, {Psychogyios}, {Tsai}, \& {Wu}}]{Assef+20}
{Assef}, R.~J., {Brightman}, M., {Walton}, D.~J., {et~al.} 2020, \apj, 897, 112, \dodoi{10.3847/1538-4357/ab9814}

\bibitem[{{Balokovi{\'c}} {et~al.}(2018){Balokovi{\'c}}, {Brightman}, {Harrison}, {Comastri}, {Ricci}, {Buchner}, {Gandhi}, {Farrah}, \& {Stern}}]{Balokovic+18}
{Balokovi{\'c}}, M., {Brightman}, M., {Harrison}, F.~A., {et~al.} 2018, \apj, 854, 42, \dodoi{10.3847/1538-4357/aaa7eb}

\bibitem[{{Birchall} {et~al.}(2023){Birchall}, {Watson}, {Aird}, \& {Starling}}]{Birchall+23}
{Birchall}, K.~L., {Watson}, M.~G., {Aird}, J., \& {Starling}, R.~L.~C. 2023, \mnras, 523, 4756, \dodoi{10.1093/mnras/stad1723}

\bibitem[{{Boquien} {et~al.}(2019){Boquien}, {Burgarella}, {Roehlly}, {Buat}, {Ciesla}, {Corre}, {Inoue}, \& {Salas}}]{Boquien+19}
{Boquien}, M., {Burgarella}, D., {Roehlly}, Y., {et~al.} 2019, \aap, 622, A103, \dodoi{10.1051/0004-6361/201834156}

\bibitem[{{Brammer} {et~al.}(2008){Brammer}, {van Dokkum}, \& {Coppi}}]{Brammer+08}
{Brammer}, G.~B., {van Dokkum}, P.~G., \& {Coppi}, P. 2008, \apj, 686, 1503, \dodoi{10.1086/591786}

\bibitem[{Brandt {et~al.}(2018)Brandt, Ni, Yang, {et~al.}}]{Brandt+18_ddfs}
Brandt, W.~N., Ni, Q., Yang, G., {et~al.} 2018,  arXiv, \dodoi{10.48550/ARXIV.1811.06542}

\bibitem[{{Brusa} {et~al.}(2015){Brusa}, {Bongiorno}, {Cresci}, {Perna}, {Marconi}, {Mainieri}, {Maiolino}, {Salvato}, {Lusso}, {Santini}, {Comastri}, {Fiore}, {Gilli}, {La Franca}, {Lanzuisi}, {Lutz}, {Merloni}, {Mignoli}, {Onori}, {Piconcelli}, {Rosario}, {Vignali}, \& {Zamorani}}]{Brusa+15}
{Brusa}, M., {Bongiorno}, A., {Cresci}, G., {et~al.} 2015, \mnras, 446, 2394, \dodoi{10.1093/mnras/stu2117}

\bibitem[{{Bruzual} \& {Charlot}(2003)}]{Bruzual+03}
{Bruzual}, G., \& {Charlot}, S. 2003, \mnras, 344, 1000, \dodoi{10.1046/j.1365-8711.2003.06897.x}

\bibitem[{{Calzetti} {et~al.}(2000){Calzetti}, {Armus}, {Bohlin}, {Kinney}, {Koornneef}, \& {Storchi-Bergmann}}]{Calzetti+00}
{Calzetti}, D., {Armus}, L., {Bohlin}, R.~C., {et~al.} 2000, \apj, 533, 682, \dodoi{10.1086/308692}

\bibitem[{{Chabrier}(2003)}]{Chabrier+03}
{Chabrier}, G. 2003, \apjl, 586, L133, \dodoi{10.1086/374879}

\bibitem[{{Chen} {et~al.}(2018){Chen}, {Brandt}, {Luo}, {Ranalli}, {Yang}, {Alexander}, {Bauer}, {Kelson}, {Lacy}, {Nyland}, {Tozzi}, {Vito}, {Cirasuolo}, {Gilli}, {Jarvis}, {Lehmer}, {Paolillo}, {Schneider}, {Shemmer}, {Smail}, {Sun}, {Tanaka}, {Vaccari}, {Vignali}, {Xue}, {Banerji}, {Chow}, {H{\"a}u{\ss}ler}, {Norris}, {Silverman}, \& {Trump}}]{Chen+18}
{Chen}, C. T.~J., {Brandt}, W.~N., {Luo}, B., {et~al.} 2018, \mnras, 478, 2132, \dodoi{10.1093/mnras/sty1036}

\bibitem[{{Ciesla} {et~al.}(2016){Ciesla}, {Boselli}, {Elbaz}, {Boissier}, {Buat}, {Charmandaris}, {Schreiber}, {B{\'e}thermin}, {Baes}, {Boquien}, {De Looze}, {Fern{\'a}ndez-Ontiveros}, {Pappalardo}, {Spinoglio}, \& {Viaene}}]{Ciesla+16}
{Ciesla}, L., {Boselli}, A., {Elbaz}, D., {et~al.} 2016, \aap, 585, A43, \dodoi{10.1051/0004-6361/201527107}

\bibitem[{{Corral} {et~al.}(2016){Corral}, {Georgantopoulos}, {Comastri}, {Ranalli}, {Akylas}, {Salvato}, {Lanzuisi}, {Vignali}, \& {Koutoulidis}}]{Corral+16}
{Corral}, A., {Georgantopoulos}, I., {Comastri}, A., {et~al.} 2016, \aap, 592, A109, \dodoi{10.1051/0004-6361/201527624}

\bibitem[{{Cristello} {et~al.}(2024){Cristello}, {Zou}, {Brandt}, {Chen}, {Leja}, {Ni}, \& {Yang}}]{Cristello+24}
{Cristello}, N., {Zou}, F., {Brandt}, W.~N., {et~al.} 2024, \apj, 962, 156, \dodoi{10.3847/1538-4357/ad2177}

\bibitem[{{Dey} {et~al.}(2008){Dey}, {Soifer}, {Desai}, {Brand}, {Le Floc'h}, {Brown}, {Jannuzi}, {Armus}, {Bussmann}, {Brodwin}, {Bian}, {Eisenhardt}, {Higdon}, {Weedman}, \& {Willner}}]{Dey+08}
{Dey}, A., {Soifer}, B.~T., {Desai}, V., {et~al.} 2008, \apj, 677, 943, \dodoi{10.1086/529516}

\bibitem[{{Doe} {et~al.}(2007){Doe}, {Nguyen}, {Stawarz}, {Refsdal}, {Siemiginowska}, {Burke}, {Evans}, {Evans}, {McDowell}, {Houck}, \& {Nowak}}]{Doe+07}
{Doe}, S., {Nguyen}, D., {Stawarz}, C., {et~al.} 2007, in Astronomical Society of the Pacific Conference Series, Vol. 376, Astronomical Data Analysis Software and Systems XVI, ed. R.~A. {Shaw}, F.~{Hill}, \& D.~J. {Bell}, 543

\bibitem[{{Draine} \& {Li}(2007)}]{Draine+07}
{Draine}, B.~T., \& {Li}, A. 2007, \apj, 657, 810, \dodoi{10.1086/511055}

\bibitem[{{Draine} {et~al.}(2014){Draine}, {Aniano}, {Krause}, {Groves}, {Sandstrom}, {Braun}, {Leroy}, {Klaas}, {Linz}, {Rix}, {Schinnerer}, {Schmiedeke}, \& {Walter}}]{Draine+14}
{Draine}, B.~T., {Aniano}, G., {Krause}, O., {et~al.} 2014, \apj, 780, 172, \dodoi{10.1088/0004-637X/780/2/172}

\bibitem[{{Gaskin} {et~al.}(2019){Gaskin}, {Swartz}, {Vikhlinin}, {{\"O}zel}, {Gelmis}, {Arenberg}, {Bandler}, {Bautz}, {Civitani}, {Dominguez}, {Eckart}, {Falcone}, {Figueroa-Feliciano}, {Freeman}, {G{\"u}nther}, {Havey}, {Heilmann}, {Kilaru}, {Kraft}, {McCarley}, {McEntaffer}, {Pareschi}, {Purcell}, {Reid}, {Schattenburg}, {Schwartz}, {Schwartz}, {Tananbaum}, {Tremblay}, {Zhang}, \& {Zuhone}}]{Gaskin+19}
{Gaskin}, J.~A., {Swartz}, D.~A., {Vikhlinin}, A., {et~al.} 2019, Journal of Astronomical Telescopes, Instruments, and Systems, 5, 021001, \dodoi{10.1117/1.JATIS.5.2.021001}

\bibitem[{{Glikman} {et~al.}(2012){Glikman}, {Urrutia}, {Lacy}, {Djorgovski}, {Mahabal}, {Myers}, {Ross}, {Petitjean}, {Ge}, {Schneider}, \& {York}}]{Glikman+12}
{Glikman}, E., {Urrutia}, T., {Lacy}, M., {et~al.} 2012, \apj, 757, 51, \dodoi{10.1088/0004-637X/757/1/51}

\bibitem[{{Goulding} {et~al.}(2018){Goulding}, {Zakamska}, {Alexandroff}, {Assef}, {Banerji}, {Hamann}, {Wylezalek}, {Brandt}, {Greene}, {Lansbury}, {P{\^a}ris}, {Richards}, {Stern}, \& {Strauss}}]{Goulding+18}
{Goulding}, A.~D., {Zakamska}, N.~L., {Alexandroff}, R.~M., {et~al.} 2018, \apj, 856, 4, \dodoi{10.3847/1538-4357/aab040}

\bibitem[{{HI4PI Collaboration} {et~al.}(2016){HI4PI Collaboration}, {Ben Bekhti}, {Fl{\"o}er}, {Keller}, {Kerp}, {Lenz}, {Winkel}, {Bailin}, {Calabretta}, {Dedes}, {Ford}, {Gibson}, {Haud}, {Janowiecki}, {Kalberla}, {Lockman}, {McClure-Griffiths}, {Murphy}, {Nakanishi}, {Pisano}, \& {Staveley-Smith}}]{heasarc}
{HI4PI Collaboration}, {Ben Bekhti}, N., {Fl{\"o}er}, L., {et~al.} 2016, \aap, 594, A116, \dodoi{10.1051/0004-6361/201629178}

\bibitem[{{Hopkins} {et~al.}(2006){Hopkins}, {Hernquist}, {Cox}, {Di Matteo}, {Robertson}, \& {Springel}}]{Hopkins+06}
{Hopkins}, P.~F., {Hernquist}, L., {Cox}, T.~J., {et~al.} 2006, \apjs, 163, 1, \dodoi{10.1086/499298}

\bibitem[{{Hopkins} {et~al.}(2008){Hopkins}, {Hernquist}, {Cox}, \& {Kere{\v{s}}}}]{Hopkins+08}
{Hopkins}, P.~F., {Hernquist}, L., {Cox}, T.~J., \& {Kere{\v{s}}}, D. 2008, \apjs, 175, 356, \dodoi{10.1086/524362}

\bibitem[{{Ishibashi} {et~al.}(2018){Ishibashi}, {Fabian}, {Ricci}, \& {Celotti}}]{Ishibashi+18}
{Ishibashi}, W., {Fabian}, A.~C., {Ricci}, C., \& {Celotti}, A. 2018, \mnras, 479, 3335, \dodoi{10.1093/mnras/sty1620}

\bibitem[{{Ishihara} {et~al.}(2010){Ishihara}, {Onaka}, {Kataza}, {Salama}, {Alfageme}, {Cassatella}, {Cox}, {Garc{\'\i}a-Lario}, {Stephenson}, {Cohen}, {Fujishiro}, {Fujiwara}, {Hasegawa}, {Ita}, {Kim}, {Matsuhara}, {Murakami}, {M{\"u}ller}, {Nakagawa}, {Ohyama}, {Oyabu}, {Pyo}, {Sakon}, {Shibai}, {Takita}, {Tanab{\'e}}, {Uemizu}, {Ueno}, {Usui}, {Wada}, {Watarai}, {Yamamura}, \& {Yamauchi}}]{Ishihara+10}
{Ishihara}, D., {Onaka}, T., {Kataza}, H., {et~al.} 2010, \aap, 514, A1, \dodoi{10.1051/0004-6361/200913811}

\bibitem[{{Ivezi{\'c}} {et~al.}(2019){Ivezi{\'c}}, {Kahn}, {Tyson}, {Abel}, {Acosta}, {Allsman}, {Alonso}, {AlSayyad}, {Anderson}, {Andrew}, {Angel}, {Angeli}, {Ansari}, {Antilogus}, {Araujo}, {Armstrong}, {Arndt}, {Astier}, {Aubourg}, {Auza}, {Axelrod}, {Bard}, {Barr}, {Barrau}, {Bartlett}, {Bauer}, {Bauman}, {Baumont}, {Bechtol}, {Bechtol}, {Becker}, {Becla}, {Beldica}, {Bellavia}, {Bianco}, {Biswas}, {Blanc}, {Blazek}, {Blandford}, {Bloom}, {Bogart}, {Bond}, {Booth}, {Borgland}, {Borne}, {Bosch}, {Boutigny}, {Brackett}, {Bradshaw}, {Brandt}, {Brown}, {Bullock}, {Burchat}, {Burke}, {Cagnoli}, {Calabrese}, {Callahan}, {Callen}, {Carlin}, {Carlson}, {Chandrasekharan}, {Charles-Emerson}, {Chesley}, {Cheu}, {Chiang}, {Chiang}, {Chirino}, {Chow}, {Ciardi}, {Claver}, {Cohen-Tanugi}, {Cockrum}, {Coles}, {Connolly}, {Cook}, {Cooray}, {Covey}, {Cribbs}, {Cui}, {Cutri}, {Daly}, {Daniel}, {Daruich}, {Daubard}, {Daues}, {Dawson}, {Delgado}, {Dellapenna}, {de Peyster}, {de Val-Borro}, {Digel}, {Doherty}, {Dubois},
  {Dubois-Felsmann}, {Durech}, {Economou}, {Eifler}, {Eracleous}, {Emmons}, {Fausti Neto}, {Ferguson}, {Figueroa}, {Fisher-Levine}, {Focke}, {Foss}, {Frank}, {Freemon}, {Gangler}, {Gawiser}, {Geary}, {Gee}, {Geha}, {Gessner}, {Gibson}, {Gilmore}, {Glanzman}, {Glick}, {Goldina}, {Goldstein}, {Goodenow}, {Graham}, {Gressler}, {Gris}, {Guy}, {Guyonnet}, {Haller}, {Harris}, {Hascall}, {Haupt}, {Hernandez}, {Herrmann}, {Hileman}, {Hoblitt}, {Hodgson}, {Hogan}, {Howard}, {Huang}, {Huffer}, {Ingraham}, {Innes}, {Jacoby}, {Jain}, {Jammes}, {Jee}, {Jenness}, {Jernigan}, {Jevremovi{\'c}}, {Johns}, {Johnson}, {Johnson}, {Jones}, {Juramy-Gilles}, {Juri{\'c}}, {Kalirai}, {Kallivayalil}, {Kalmbach}, {Kantor}, {Karst}, {Kasliwal}, {Kelly}, {Kessler}, {Kinnison}, {Kirkby}, {Knox}, {Kotov}, {Krabbendam}, {Krughoff}, {Kub{\'a}nek}, {Kuczewski}, {Kulkarni}, {Ku}, {Kurita}, {Lage}, {Lambert}, {Lange}, {Langton}, {Le Guillou}, {Levine}, {Liang}, {Lim}, {Lintott}, {Long}, {Lopez}, {Lotz}, {Lupton}, {Lust}, {MacArthur}, {Mahabal},
  {Mandelbaum}, {Markiewicz}, {Marsh}, {Marshall}, {Marshall}, {May}, {McKercher}, {McQueen}, {Meyers}, {Migliore}, {Miller}, {Mills}, {Miraval}, {Moeyens}, {Moolekamp}, {Monet}, {Moniez}, {Monkewitz}, {Montgomery}, {Morrison}, {Mueller}, {Muller}, {Mu{\~n}oz Arancibia}, {Neill}, {Newbry}, {Nief}, {Nomerotski}, {Nordby}, {O'Connor}, {Oliver}, {Olivier}, {Olsen}, {O'Mullane}, {Ortiz}, {Osier}, {Owen}, {Pain}, {Palecek}, {Parejko}, {Parsons}, {Pease}, {Peterson}, {Peterson}, {Petravick}, {Libby Petrick}, {Petry}, {Pierfederici}, {Pietrowicz}, {Pike}, {Pinto}, {Plante}, {Plate}, {Plutchak}, {Price}, {Prouza}, {Radeka}, {Rajagopal}, {Rasmussen}, {Regnault}, {Reil}, {Reiss}, {Reuter}, {Ridgway}, {Riot}, {Ritz}, {Robinson}, {Roby}, {Roodman}, {Rosing}, {Roucelle}, {Rumore}, {Russo}, {Saha}, {Sassolas}, {Schalk}, {Schellart}, {Schindler}, {Schmidt}, {Schneider}, {Schneider}, {Schoening}, {Schumacher}, {Schwamb}, {Sebag}, {Selvy}, {Sembroski}, {Seppala}, {Serio}, {Serrano}, {Shaw}, {Shipsey}, {Sick}, {Silvestri},
  {Slater}, {Smith}, {Smith}, {Sobhani}, {Soldahl}, {Storrie-Lombardi}, {Stover}, {Strauss}, {Street}, {Stubbs}, {Sullivan}, {Sweeney}, {Swinbank}, {Szalay}, {Takacs}, {Tether}, {Thaler}, {Thayer}, {Thomas}, {Thornton}, {Thukral}, {Tice}, {Trilling}, {Turri}, {Van Berg}, {Vanden Berk}, {Vetter}, {Virieux}, {Vucina}, {Wahl}, {Walkowicz}, {Walsh}, {Walter}, {Wang}, {Wang}, {Warner}, {Wiecha}, {Willman}, {Winters}, {Wittman}, {Wolff}, {Wood-Vasey}, {Wu}, {Xin}, {Yoachim}, \& {Zhan}}]{Ivesic+19_lsst}
{Ivezi{\'c}}, {\v{Z}}., {Kahn}, S.~M., {Tyson}, J.~A., {et~al.} 2019, \apj, 873, 111, \dodoi{10.3847/1538-4357/ab042c}

\bibitem[{{Jansen} {et~al.}(2001){Jansen}, {Lumb}, {Altieri}, {Clavel}, {Ehle}, {Erd}, {Gabriel}, {Guainazzi}, {Gondoin}, {Much}, {Munoz}, {Santos}, {Schartel}, {Texier}, \& {Vacanti}}]{Jansen+01}
{Jansen}, F., {Lumb}, D., {Altieri}, B., {et~al.} 2001, \aap, 365, L1, \dodoi{10.1051/0004-6361:20000036}

\bibitem[{{Johnson} {et~al.}(2021){Johnson}, {Leja}, {Conroy}, \& {Speagle}}]{Johnson+21}
{Johnson}, B.~D., {Leja}, J., {Conroy}, C., \& {Speagle}, J.~S. 2021, \apjs, 254, 22, \dodoi{10.3847/1538-4365/abef67}

\bibitem[{{Kawada} {et~al.}(2007){Kawada}, {Baba}, {Barthel}, {Clements}, {Cohen}, {Doi}, {Figueredo}, {Fujiwara}, {Goto}, {Hasegawa}, {Hibi}, {Hirao}, {Hiromoto}, {Jeong}, {Kaneda}, {Kawai}, {Kawamura}, {Kester}, {Kii}, {Kobayashi}, {Kwon}, {Lee}, {Makiuti}, {Matsuo}, {Matsuura}, {M{\"u}ller}, {Murakami}, {Nagata}, {Nakagawa}, {Narita}, {Noda}, {Oh}, {Okada}, {Okuda}, {Oliver}, {Ootsubo}, {Pak}, {Park}, {Pearson}, {Rowan-Robinson}, {Saito}, {Salama}, {Sato}, {Savage}, {Serjeant}, {Shibai}, {Shirahata}, {Sohn}, {Suzuki}, {Takagi}, {Takahashi}, {Thomson}, {Usui}, {Verdugo}, {Watabe}, {White}, {Wang}, {Yamamura}, {Yamauchi}, \& {Yasuda}}]{Kawada+07}
{Kawada}, M., {Baba}, H., {Barthel}, P.~D., {et~al.} 2007, \pasj, 59, S389, \dodoi{10.1093/pasj/59.sp2.S389}

\bibitem[{{Kayal} \& {Singh}(2024)}]{Kayal+24}
{Kayal}, A., \& {Singh}, V. 2024, \mnras, 531, 830, \dodoi{10.1093/mnras/stae1191}

\bibitem[{{Lansbury} {et~al.}(2020){Lansbury}, {Banerji}, {Fabian}, \& {Temple}}]{Lansbury+20}
{Lansbury}, G.~B., {Banerji}, M., {Fabian}, A.~C., \& {Temple}, M.~J. 2020, \mnras, 495, 2652, \dodoi{10.1093/mnras/staa1220}

\bibitem[{{Lanzuisi} {et~al.}(2009){Lanzuisi}, {Piconcelli}, {Fiore}, {Feruglio}, {Vignali}, {Salvato}, \& {Gruppioni}}]{Lanzuisi+09}
{Lanzuisi}, G., {Piconcelli}, E., {Fiore}, F., {et~al.} 2009, \aap, 498, 67, \dodoi{10.1051/0004-6361/200811282}

\bibitem[{{Lanzuisi} {et~al.}(2015){Lanzuisi}, {Ranalli}, {Georgantopoulos}, {Georgakakis}, {Delvecchio}, {Akylas}, {Berta}, {Bongiorno}, {Brusa}, {Cappelluti}, {Civano}, {Comastri}, {Gilli}, {Gruppioni}, {Hasinger}, {Iwasawa}, {Koekemoer}, {Lusso}, {Marchesi}, {Mainieri}, {Merloni}, {Mignoli}, {Piconcelli}, {Pozzi}, {Rosario}, {Salvato}, {Silverman}, {Trakhtenbrot}, {Vignali}, \& {Zamorani}}]{Lanzuisi+15}
{Lanzuisi}, G., {Ranalli}, P., {Georgantopoulos}, I., {et~al.} 2015, \aap, 573, A137, \dodoi{10.1051/0004-6361/201424924}

\bibitem[{{Leja} {et~al.}(2019){Leja}, {Carnall}, {Johnson}, {Conroy}, \& {Speagle}}]{Leja+19a}
{Leja}, J., {Carnall}, A.~C., {Johnson}, B.~D., {Conroy}, C., \& {Speagle}, J.~S. 2019, \apj, 876, 3, \dodoi{10.3847/1538-4357/ab133c}

\bibitem[{{Leja} {et~al.}(2017){Leja}, {Johnson}, {Conroy}, {van Dokkum}, \& {Byler}}]{Leja+17}
{Leja}, J., {Johnson}, B.~D., {Conroy}, C., {van Dokkum}, P.~G., \& {Byler}, N. 2017, \apj, 837, 170, \dodoi{10.3847/1538-4357/aa5ffe}

\bibitem[{{Li} {et~al.}(2019){Li}, {Xue}, {Sun}, {Liu}, {Vito}, {Brandt}, {Hughes}, {Yang}, {Tozzi}, {Zhu}, {Zheng}, {Luo}, {Chen}, {Vignali}, {Gilli}, \& {Shu}}]{Li+19}
{Li}, J., {Xue}, Y., {Sun}, M., {et~al.} 2019, \apj, 877, 5, \dodoi{10.3847/1538-4357/ab184b}

\bibitem[{{Magdziarz} \& {Zdziarski}(1995)}]{Magdziarz+95}
{Magdziarz}, P., \& {Zdziarski}, A.~A. 1995, \mnras, 273, 837, \dodoi{10.1093/mnras/273.3.837}

\bibitem[{{Martocchia} {et~al.}(2017){Martocchia}, {Piconcelli}, {Zappacosta}, {Duras}, {Vietri}, {Vignali}, {Bianchi}, {Bischetti}, {Bongiorno}, {Brusa}, {Lanzuisi}, {Marconi}, {Mathur}, {Miniutti}, {Nicastro}, {Bruni}, \& {Fiore}}]{Martocchia+17}
{Martocchia}, S., {Piconcelli}, E., {Zappacosta}, L., {et~al.} 2017, \aap, 608, A51, \dodoi{10.1051/0004-6361/201731314}

\bibitem[{{Miller} {et~al.}(2011){Miller}, {Brandt}, {Schneider}, {Gibson}, {Steffen}, \& {Wu}}]{Miller+11}
{Miller}, B.~P., {Brandt}, W.~N., {Schneider}, D.~P., {et~al.} 2011, \apj, 726, 20, \dodoi{10.1088/0004-637X/726/1/20}

\bibitem[{{Mountrichas} {et~al.}(2021){Mountrichas}, {Buat}, {Yang}, {Boquien}, {Burgarella}, {Ciesla}, {Malek}, \& {Shirley}}]{Mountrichas+21}
{Mountrichas}, G., {Buat}, V., {Yang}, G., {et~al.} 2021, \aap, 653, A74, \dodoi{10.1051/0004-6361/202140630}

\bibitem[{{Mountrichas} {et~al.}(2024){Mountrichas}, {Ruiz}, {Georgantopoulos}, {Pouliasis}, {Akylas}, \& {Drigga}}]{Mountrichas+24b}
{Mountrichas}, G., {Ruiz}, A., {Georgantopoulos}, I., {et~al.} 2024, \aap, 688, A79, \dodoi{10.1051/0004-6361/202449601}

\bibitem[{{Mountrichas} {et~al.}(2017){Mountrichas}, {Georgantopoulos}, {Secrest}, {Ordov{\'a}s-Pascual}, {Corral}, {Akylas}, {Mateos}, {Carrera}, \& {Batziou}}]{Mountrichas+17}
{Mountrichas}, G., {Georgantopoulos}, I., {Secrest}, N.~J., {et~al.} 2017, \mnras, 468, 3042, \dodoi{10.1093/mnras/stx572}

\bibitem[{{Mullaney} {et~al.}(2015){Mullaney}, {Alexander}, {Aird}, {Bernhard}, {Daddi}, {Del Moro}, {Dickinson}, {Elbaz}, {Harrison}, {Juneau}, {Liu}, {Pannella}, {Rosario}, {Santini}, {Sargent}, {Schreiber}, {Simpson}, \& {Stanley}}]{Mullaney+15}
{Mullaney}, J.~R., {Alexander}, D.~M., {Aird}, J., {et~al.} 2015, \mnras, 453, L83, \dodoi{10.1093/mnrasl/slv110}

\bibitem[{{Nandra} {et~al.}(2013){Nandra}, {Barret}, {Barcons}, {Fabian}, {den Herder}, {Piro}, {Watson}, {Adami}, {Aird}, {Afonso}, {Alexander}, {Argiroffi}, {Amati}, {Arnaud}, {Atteia}, {Audard}, {Badenes}, {Ballet}, {Ballo}, {Bamba}, {Bhardwaj}, {Stefano Battistelli}, {Becker}, {De Becker}, {Behar}, {Bianchi}, {Biffi}, {B{\^\i}rzan}, {Bocchino}, {Bogdanov}, {Boirin}, {Boller}, {Borgani}, {Borm}, {Bouch{\'e}}, {Bourdin}, {Bower}, {Braito}, {Branchini}, {Branduardi-Raymont}, {Bregman}, {Brenneman}, {Brightman}, {Br{\"u}ggen}, {Buchner}, {Bulbul}, {Brusa}, {Bursa}, {Caccianiga}, {Cackett}, {Campana}, {Cappelluti}, {Cappi}, {Carrera}, {Ceballos}, {Christensen}, {Chu}, {Churazov}, {Clerc}, {Corbel}, {Corral}, {Comastri}, {Costantini}, {Croston}, {Dadina}, {D'Ai}, {Decourchelle}, {Della Ceca}, {Dennerl}, {Dolag}, {Done}, {Dovciak}, {Drake}, {Eckert}, {Edge}, {Ettori}, {Ezoe}, {Feigelson}, {Fender}, {Feruglio}, {Finoguenov}, {Fiore}, {Galeazzi}, {Gallagher}, {Gandhi}, {Gaspari}, {Gastaldello}, {Georgakakis},
  {Georgantopoulos}, {Gilfanov}, {Gitti}, {Gladstone}, {Goosmann}, {Gosset}, {Grosso}, {Guedel}, {Guerrero}, {Haberl}, {Hardcastle}, {Heinz}, {Alonso Herrero}, {Herv{\'e}}, {Holmstrom}, {Iwasawa}, {Jonker}, {Kaastra}, {Kara}, {Karas}, {Kastner}, {King}, {Kosenko}, {Koutroumpa}, {Kraft}, {Kreykenbohm}, {Lallement}, {Lanzuisi}, {Lee}, {Lemoine-Goumard}, {Lobban}, {Lodato}, {Lovisari}, {Lotti}, {McCharthy}, {McNamara}, {Maggio}, {Maiolino}, {De Marco}, {de Martino}, {Mateos}, {Matt}, {Maughan}, {Mazzotta}, {Mendez}, {Merloni}, {Micela}, {Miceli}, {Mignani}, {Miller}, {Miniutti}, {Molendi}, {Montez}, {Moretti}, {Motch}, {Naz{\'e}}, {Nevalainen}, {Nicastro}, {Nulsen}, {Ohashi}, {O'Brien}, {Osborne}, {Oskinova}, {Pacaud}, {Paerels}, {Page}, {Papadakis}, {Pareschi}, {Petre}, {Petrucci}, {Piconcelli}, {Pillitteri}, {Pinto}, {de Plaa}, {Pointecouteau}, {Ponman}, {Ponti}, {Porquet}, {Pounds}, {Pratt}, {Predehl}, {Proga}, {Psaltis}, {Rafferty}, {Ramos-Ceja}, {Ranalli}, {Rasia}, {Rau}, {Rauw}, {Rea}, {Read}, {Reeves},
  {Reiprich}, {Renaud}, {Reynolds}, {Risaliti}, {Rodriguez}, {Rodriguez Hidalgo}, {Roncarelli}, {Rosario}, {Rossetti}, {Rozanska}, {Rovilos}, {Salvaterra}, {Salvato}, {Di Salvo}, {Sanders}, {Sanz-Forcada}, {Schawinski}, {Schaye}, {Schwope}, {Sciortino}, {Severgnini}, {Shankar}, {Sijacki}, {Sim}, {Schmid}, {Smith}, {Steiner}, {Stelzer}, {Stewart}, {Strohmayer}, {Str{\"u}der}, {Sun}, {Takei}, {Tatischeff}, {Tiengo}, {Tombesi}, {Trinchieri}, {Tsuru}, {Ud-Doula}, {Ursino}, {Valencic}, {Vanzella}, {Vaughan}, {Vignali}, {Vink}, {Vito}, {Volonteri}, {Wang}, {Webb}, {Willingale}, {Wilms}, {Wise}, {Worrall}, {Young}, {Zampieri}, {In't Zand}, {Zane}, {Zezas}, {Zhang}, \& {Zhuravleva}}]{Nandra+13}
{Nandra}, K., {Barret}, D., {Barcons}, X., {et~al.} 2013, arXiv e-prints, arXiv:1306.2307, \dodoi{10.48550/arXiv.1306.2307}

\bibitem[{{Netzer}(2015)}]{Netzer+15}
{Netzer}, H. 2015, \araa, 53, 365, \dodoi{10.1146/annurev-astro-082214-122302}

\bibitem[{{Ni} {et~al.}(2021){Ni}, {Brandt}, {Chen}, {Luo}, {Nyland}, {Yang}, {Zou}, {Aird}, {Alexander}, {Bauer}, {Lacy}, {Lehmer}, {Mallick}, {Salvato}, {Schneider}, {Tozzi}, {Traulsen}, {Vaccari}, {Vignali}, {Vito}, {Xue}, {Banerji}, {Chow}, {Comastri}, {Del Moro}, {Gilli}, {Mullaney}, {Paolillo}, {Schwope}, {Shemmer}, {Sun}, {Timlin}, \& {Trump}}]{Ni+21_xmmservs}
{Ni}, Q., {Brandt}, W.~N., {Chen}, C.-T., {et~al.} 2021, \apjs, 256, 21, \dodoi{10.3847/1538-4365/ac0dc6}

\bibitem[{{Noboriguchi} {et~al.}(2023){Noboriguchi}, {Inoue}, {Nagao}, {Toba}, \& {Misawa}}]{Nob+23}
{Noboriguchi}, A., {Inoue}, A.~K., {Nagao}, T., {Toba}, Y., \& {Misawa}, T. 2023, \apjl, 959, L14, \dodoi{10.3847/2041-8213/ad0e00}

\bibitem[{{Noboriguchi} {et~al.}(2022){Noboriguchi}, {Nagao}, {Toba}, {Ichikawa}, {Kajisawa}, {Kato}, {Kawaguchi}, {Matsuhara}, {Matsuoka}, {Onishi}, {Onoue}, {Tamada}, {Terao}, {Terashima}, {Ueda}, \& {Yamashita}}]{Nob+22}
{Noboriguchi}, A., {Nagao}, T., {Toba}, Y., {et~al.} 2022, \apj, 941, 195, \dodoi{10.3847/1538-4357/aca403}

\bibitem[{{Oke} \& {Gunn}(1983)}]{Oke+83}
{Oke}, J.~B., \& {Gunn}, J.~E. 1983, \apj, 266, 713, \dodoi{10.1086/160817}

\bibitem[{{Piconcelli} {et~al.}(2015){Piconcelli}, {Vignali}, {Bianchi}, {Zappacosta}, {Fritz}, {Lanzuisi}, {Miniutti}, {Bongiorno}, {Feruglio}, {Fiore}, \& {Maiolino}}]{Piconcelli+15}
{Piconcelli}, E., {Vignali}, C., {Bianchi}, S., {et~al.} 2015, \aap, 574, L9, \dodoi{10.1051/0004-6361/201425324}

\bibitem[{{Popesso} {et~al.}(2023){Popesso}, {Concas}, {Cresci}, {Belli}, {Rodighiero}, {Inami}, {Dickinson}, {Ilbert}, {Pannella}, \& {Elbaz}}]{Popesso+23}
{Popesso}, P., {Concas}, A., {Cresci}, G., {et~al.} 2023, \mnras, 519, 1526, \dodoi{10.1093/mnras/stac3214}

\bibitem[{{Prevot} {et~al.}(1984){Prevot}, {Lequeux}, {Maurice}, {Prevot}, \& {Rocca-Volmerange}}]{Prevot+84}
{Prevot}, M.~L., {Lequeux}, J., {Maurice}, E., {Prevot}, L., \& {Rocca-Volmerange}, B. 1984, \aap, 132, 389

\bibitem[{{Ranalli} {et~al.}(2013){Ranalli}, {Comastri}, {Vignali}, {Carrera}, {Cappelluti}, {Gilli}, {Puccetti}, {Brandt}, {Brunner}, {Brusa}, {Georgantopoulos}, {Iwasawa}, \& {Mainieri}}]{Ranalli+13}
{Ranalli}, P., {Comastri}, A., {Vignali}, C., {et~al.} 2013, \aap, 555, A42, \dodoi{10.1051/0004-6361/201321211}

\bibitem[{{Reines} \& {Volonteri}(2015)}]{Reines+15}
{Reines}, A.~E., \& {Volonteri}, M. 2015, \apj, 813, 82, \dodoi{10.1088/0004-637X/813/2/82}

\bibitem[{{Ricci} {et~al.}(2017){Ricci}, {Assef}, {Stern}, {Nikutta}, {Alexander}, {Asmus}, {Ballantyne}, {Bauer}, {Blain}, {Boggs}, {Boorman}, {Brandt}, {Brightman}, {Chang}, {Chen}, {Christensen}, {Comastri}, {Craig}, {D{\'\i}az-Santos}, {Eisenhardt}, {Farrah}, {Gandhi}, {Hailey}, {Harrison}, {Jun}, {Koss}, {LaMassa}, {Lansbury}, {Markwardt}, {Stalevski}, {Stanley}, {Treister}, {Tsai}, {Walton}, {Wu}, {Zappacosta}, \& {Zhang}}]{Ricci+17}
{Ricci}, C., {Assef}, R.~J., {Stern}, D., {et~al.} 2017, \apj, 835, 105, \dodoi{10.3847/1538-4357/835/1/105}

\bibitem[{{Schartmann} {et~al.}(2005){Schartmann}, {Meisenheimer}, {Camenzind}, {Wolf}, \& {Henning}}]{Schartmann+05}
{Schartmann}, M., {Meisenheimer}, K., {Camenzind}, M., {Wolf}, S., \& {Henning}, T. 2005, \aap, 437, 861, \dodoi{10.1051/0004-6361:20042363}

\bibitem[{{Stalevski} {et~al.}(2012){Stalevski}, {Fritz}, {Baes}, {Nakos}, \& {Popovi{\'c}}}]{Stalevski+12}
{Stalevski}, M., {Fritz}, J., {Baes}, M., {Nakos}, T., \& {Popovi{\'c}}, L.~{\v{C}}. 2012, \mnras, 420, 2756, \dodoi{10.1111/j.1365-2966.2011.19775.x}

\bibitem[{{Stalevski} {et~al.}(2016){Stalevski}, {Ricci}, {Ueda}, {Lira}, {Fritz}, \& {Baes}}]{Stalevski+16}
{Stalevski}, M., {Ricci}, C., {Ueda}, Y., {et~al.} 2016, \mnras, 458, 2288, \dodoi{10.1093/mnras/stw444}

\bibitem[{{Stern}(2015)}]{Stern+15}
{Stern}, D. 2015, \apj, 807, 129, \dodoi{10.1088/0004-637X/807/2/129}

\bibitem[{{Stern} {et~al.}(2014){Stern}, {Lansbury}, {Assef}, {Brandt}, {Alexander}, {Ballantyne}, {Balokovi{\'c}}, {Bauer}, {Benford}, {Blain}, {Boggs}, {Bridge}, {Brightman}, {Christensen}, {Comastri}, {Craig}, {Del Moro}, {Eisenhardt}, {Gandhi}, {Griffith}, {Hailey}, {Harrison}, {Hickox}, {Jarrett}, {Koss}, {Lake}, {LaMassa}, {Luo}, {Tsai}, {Urry}, {Walton}, {Wright}, {Wu}, {Yan}, \& {Zhang}}]{Stern+14}
{Stern}, D., {Lansbury}, G.~B., {Assef}, R.~J., {et~al.} 2014, \apj, 794, 102, \dodoi{10.1088/0004-637X/794/2/102}

\bibitem[{{Toba} {et~al.}(2017){Toba}, {Bae}, {Nagao}, {Woo}, {Wang}, {Wagner}, {Sun}, \& {Chang}}]{Toba+17}
{Toba}, Y., {Bae}, H.-J., {Nagao}, T., {et~al.} 2017, \apj, 850, 140, \dodoi{10.3847/1538-4357/aa918a}

\bibitem[{{Toba} \& {Nagao}(2016)}]{Toba+16}
{Toba}, Y., \& {Nagao}, T. 2016, \apj, 820, 46, \dodoi{10.3847/0004-637X/820/1/46}

\bibitem[{{Tsai} {et~al.}(2015){Tsai}, {Eisenhardt}, {Wu}, {Stern}, {Assef}, {Blain}, {Bridge}, {Benford}, {Cutri}, {Griffith}, {Jarrett}, {Lonsdale}, {Masci}, {Moustakas}, {Petty}, {Sayers}, {Stanford}, {Wright}, {Yan}, {Leisawitz}, {Liu}, {Mainzer}, {McLean}, {Padgett}, {Skrutskie}, {Gelino}, {Beichman}, \& {Juneau}}]{Tsai+15}
{Tsai}, C.-W., {Eisenhardt}, P. R.~M., {Wu}, J., {et~al.} 2015, \apj, 805, 90, \dodoi{10.1088/0004-637X/805/2/90}

\bibitem[{{Urrutia} {et~al.}(2005){Urrutia}, {Lacy}, {Gregg}, \& {Becker}}]{Urrutia+05}
{Urrutia}, T., {Lacy}, M., {Gregg}, M.~D., \& {Becker}, R.~H. 2005, \apj, 627, 75, \dodoi{10.1086/430165}

\bibitem[{{Vietri} {et~al.}(2022){Vietri}, {Garilli}, {Polletta}, {Bisogni}, {Cassar{\`a}}, {Franzetti}, {Fumana}, {Gargiulo}, {Maccagni}, {Mancini}, {Scodeggio}, {Fritz}, {Ma{\l}ek}, {Manzoni}, {Pollo}, {Siudek}, {Vergani}, {Zamorani}, \& {Zanichelli}}]{Vietri+22}
{Vietri}, G., {Garilli}, B., {Polletta}, M., {et~al.} 2022, \aap, 659, A129, \dodoi{10.1051/0004-6361/202141072}

\bibitem[{{Vito} {et~al.}(2018){Vito}, {Brandt}, {Stern}, {Assef}, {Chen}, {Brightman}, {Comastri}, {Eisenhardt}, {Garmire}, {Hickox}, {Lansbury}, {Tsai}, {Walton}, \& {Wu}}]{Vito+18}
{Vito}, F., {Brandt}, W.~N., {Stern}, D., {et~al.} 2018, \mnras, 474, 4528, \dodoi{10.1093/mnras/stx3120}

\bibitem[{{Werner} {et~al.}(2004){Werner}, {Roellig}, {Low}, {Rieke}, {Rieke}, {Hoffmann}, {Young}, {Houck}, {Brandl}, {Fazio}, {Hora}, {Gehrz}, {Helou}, {Soifer}, {Stauffer}, {Keene}, {Eisenhardt}, {Gallagher}, {Gautier}, {Irace}, {Lawrence}, {Simmons}, {Van Cleve}, {Jura}, {Wright}, \& {Cruikshank}}]{Werner+04}
{Werner}, M.~W., {Roellig}, T.~L., {Low}, F.~J., {et~al.} 2004, \apjs, 154, 1, \dodoi{10.1086/422992}

\bibitem[{{Worrall} {et~al.}(1987){Worrall}, {Giommi}, {Tananbaum}, \& {Zamorani}}]{Worrall+87}
{Worrall}, D.~M., {Giommi}, P., {Tananbaum}, H., \& {Zamorani}, G. 1987, \apj, 313, 596, \dodoi{10.1086/164999}

\bibitem[{{Wright} {et~al.}(2010){Wright}, {Eisenhardt}, {Mainzer}, {Ressler}, {Cutri}, {Jarrett}, {Kirkpatrick}, {Padgett}, {McMillan}, {Skrutskie}, {Stanford}, {Cohen}, {Walker}, {Mather}, {Leisawitz}, {Gautier}, {McLean}, {Benford}, {Lonsdale}, {Blain}, {Mendez}, {Irace}, {Duval}, {Liu}, {Royer}, {Heinrichsen}, {Howard}, {Shannon}, {Kendall}, {Walsh}, {Larsen}, {Cardon}, {Schick}, {Schwalm}, {Abid}, {Fabinsky}, {Naes}, \& {Tsai}}]{Wright+10}
{Wright}, E.~L., {Eisenhardt}, P. R.~M., {Mainzer}, A.~K., {et~al.} 2010, \aj, 140, 1868, \dodoi{10.1088/0004-6256/140/6/1868}

\bibitem[{{Wu} {et~al.}(2012){Wu}, {Tsai}, {Sayers}, {Benford}, {Bridge}, {Blain}, {Eisenhardt}, {Stern}, {Petty}, {Assef}, {Bussmann}, {Comerford}, {Cutri}, {Evans}, {Griffith}, {Jarrett}, {Lake}, {Lonsdale}, {Rho}, {Stanford}, {Weiner}, {Wright}, \& {Yan}}]{Wu+12}
{Wu}, J., {Tsai}, C.-W., {Sayers}, J., {et~al.} 2012, \apj, 756, 96, \dodoi{10.1088/0004-637X/756/1/96}

\bibitem[{{Wu} {et~al.}(2018){Wu}, {Jun}, {Assef}, {Tsai}, {Wright}, {Eisenhardt}, {Blain}, {Stern}, {D{\'\i}az-Santos}, {Denney}, {Hayden}, {Perlmutter}, {Aldering}, {Boone}, \& {Fagrelius}}]{Wu+18}
{Wu}, J., {Jun}, H.~D., {Assef}, R.~J., {et~al.} 2018, \apj, 852, 96, \dodoi{10.3847/1538-4357/aa9ff3}

\bibitem[{{Yang} {et~al.}(2020){Yang}, {Boquien}, {Buat}, {Burgarella}, {Ciesla}, {Duras}, {Stalevski}, {Brandt}, \& {Papovich}}]{Yang+20_cigale}
{Yang}, G., {Boquien}, M., {Buat}, V., {et~al.} 2020, \mnras, 491, 740, \dodoi{10.1093/mnras/stz3001}

\bibitem[{{Yang} {et~al.}(2022){Yang}, {Boquien}, {Brandt}, {Buat}, {Burgarella}, {Ciesla}, {Lehmer}, {Ma{\l}ek}, {Mountrichas}, {Papovich}, {Pons}, {Stalevski}, {Theul{\'e}}, \& {Zhu}}]{Yang+22_cigale}
{Yang}, G., {Boquien}, M., {Brandt}, W.~N., {et~al.} 2022, \apj, 927, 192, \dodoi{10.3847/1538-4357/ac4971}

\bibitem[{Yu {et~al.}(2024)Yu, Brandt, Zou, Zhu, Bauer, Cristello, Luo, Ni, Vito, \& Xue}]{Yu+24}
Yu, Z., Brandt, W.~N., Zou, F., {et~al.} 2024, arXiv e-prints, \dodoi{https://doi.org/10.48550/arXiv.2410.18190}

\bibitem[{{Zappacosta} {et~al.}(2018){Zappacosta}, {Piconcelli}, {Duras}, {Vignali}, {Valiante}, {Bianchi}, {Bongiorno}, {Fiore}, {Feruglio}, {Lanzuisi}, {Maiolino}, {Mathur}, {Miniutti}, \& {Ricci}}]{Zappacosta+18}
{Zappacosta}, L., {Piconcelli}, E., {Duras}, F., {et~al.} 2018, \aap, 618, A28, \dodoi{10.1051/0004-6361/201732557}

\bibitem[{{Zhang} \& {Yuan}(2023)}]{Zhang_R+23}
{Zhang}, R., \& {Yuan}, H. 2023, \apjs, 264, 14, \dodoi{10.3847/1538-4365/ac9dfa}

\bibitem[{{Zhu} {et~al.}(2023){Zhu}, {Brandt}, {Zou}, {Luo}, {Ni}, {Xue}, \& {Yan}}]{Zhu+23}
{Zhu}, S., {Brandt}, W.~N., {Zou}, F., {et~al.} 2023, \mnras, 522, 3506, \dodoi{10.1093/mnras/stad1178}

\bibitem[{{Zhu} {et~al.}(2020){Zhu}, {Brandt}, {Luo}, {Wu}, {Xue}, \& {Yang}}]{Zhu+20}
{Zhu}, S.~F., {Brandt}, W.~N., {Luo}, B., {et~al.} 2020, \mnras, 496, 245, \dodoi{10.1093/mnras/staa1411}

\bibitem[{{Zhu} {et~al.}(2021){Zhu}, {Timlin}, \& {Brandt}}]{Zhu+21}
{Zhu}, S.~F., {Timlin}, J.~D., \& {Brandt}, W.~N. 2021, \mnras, 505, 1954, \dodoi{10.1093/mnras/stab1406}

\bibitem[{{Zou} {et~al.}(2020){Zou}, {Brandt}, {Vito}, {Chen}, {Garmire}, {Stern}, \& {Ayubinia}}]{Zou+20}
{Zou}, F., {Brandt}, W.~N., {Vito}, F., {et~al.} 2020, \mnras, 499, 1823, \dodoi{10.1093/mnras/staa2930}

\bibitem[{{Zou} {et~al.}(2021){Zou}, {Yang}, {Brandt}, {Ni}, {Bauer}, {Covone}, {Lacy}, {Napolitano}, {Nyland}, {Paolillo}, {Radovich}, {Spavone}, \& {Vaccari}}]{Zou+21}
{Zou}, F., {Yang}, G., {Brandt}, W.~N., {et~al.} 2021, Research Notes of the American Astronomical Society, 5, 56, \dodoi{10.3847/2515-5172/abf050}

\bibitem[{{Zou} {et~al.}(2022){Zou}, {Brandt}, {Chen}, {Leja}, {Ni}, {Yan}, {Yang}, {Zhu}, {Luo}, {Nyland}, {Vito}, \& {Xue}}]{Zou+22}
{Zou}, F., {Brandt}, W.~N., {Chen}, C.-T., {et~al.} 2022, \apjs, 262, 15, \dodoi{10.3847/1538-4365/ac7bdf}

\end{thebibliography}
\bibliographystyle{aasjournal}

\end{document}